\documentclass[twocolumn,secnumarabic,amssymb, nobibnotes, aps, prd]{revtex4-1}
\usepackage{longtable}
\usepackage{graphicx}
\usepackage{amsmath,amssymb}
\usepackage{textcomp}
\usepackage[normalem]{ulem}
\usepackage{todonotes}

\begin{document} 
  
\title{Systematic calculations of reactions with exotic and stable nuclei to establish a unified theoretical approach}
      
\author{M. A. G. Alvarez$^1$, M. Rodr\'{i}guez-Gallardo$^{1,2}$, J. P. Fern\'{a}ndez-Garc\'{i}a $^{1,3}$, J. Casal$^{1,4}$ and J. A. Lay$^{1,2}$}

\affiliation{$^1$Departamento FAMN, Universidad de Sevilla, Apartado 1065, 41080 Sevilla, Spain}

\affiliation{$^2$Instituto Carlos I de F\'{i}sica Te\'{o}rica y Computacional, Universidad de Sevilla, Spain}

\affiliation{$^3$Centro Nacional de Aceleradores, Universidad de Sevilla, Junta de Andalucía-CSIC, 41092 Sevilla, Spain}

\affiliation {$^4$Dipartimento di Fisica e Astronomia ``G. Galilei'' and INFN - Sezione di Padova, Via Marzolo, 8, I-35131, Padova, Italy}

\date{\today}     
 
\begin{abstract} 

This manuscript reports on systematical optical model (OM) and continuum discretized coupled channel (CDCC) calculations applied to describe the elastic scattering angular distributions of exotic and stable nuclei projectiles on heavy targets. Our optical potential (OP) is composed by the nuclear microscopic double folding S\~ao Paulo potential (SPP), derived from the non-local nature of the interaction, and the Coulomb dipole polarization (CDP) potential, derived from the semi-classical theory of Coulomb excitation. The OP is compared to the trivial equivalent local potential (TELP), extracted from CDCC calculations. The OM and CDCC predictions corroborate each other and account for important differences in the nuclei reaction mechanisms, which are directly related to their structural properties. Thus, OM and CDCC establish a common basis for analysing or even predicting on exotic and stable nuclei reactions.
     
\end{abstract}      
      
      
\maketitle

%%%%%%%%%%%%%%%%%%%%%%%%%%%%%%%%%%%%%%%%%%%%%%%%%%%%%%%%%%%%%%%%%%%%%%%%%%%%%%%
\section{Introduction}      
\label{introduct}      
%%%%%%%%%%%%%%%%%%%%%%%%%%%%%%%%%%%%%%%%%%%%%%%%%%%%%%%%%%%%%%%%%%%%%%%%%%%%%%%
      
\indent 

Describing nuclear reactions of exotic and stable nuclei with the same theoretical approach, is one of the most challenging problems in nuclear physics. Studying reactions involving weakly bound stable nuclei is a natural step at the interface between the cases of tightly bound stable and exotic nuclei, and towards a better understanding of the latter. The structural models of these nuclei are fundamental to determine how they interact. Comparing nuclear structures, reaction mechanisms and their rates are crucial for key studies, such as explaining nuclei abundances in the cosmos \cite{RevModPhys.29.547}. In such a scenario, astrophysical applications assume optical potential (OP) approaches, as input for theoretical calculations \cite{S0375-9474(02)00756-X,1.1737132, PhysRevC.76.045802}. Folding type potentials allow to infer about nuclear structures, in a fundamental way, through taking into account realistic models for the nuclear densities \cite{PhysRevC.65.014602}.

Within such fundamental basis, other important properties can be investigated. Unlike tightly bound stable nuclei, weakly bound and exotic nuclei have one common striking characteristic: a low break-up threshold. This characteristic favors break-up and, therefore, elastic scattering flux absorption, when interacting with another nucleus. Break-up gives rise to a complex problem of three or more bodies and can be driven by different processes: core excitation, direct excitation of the projectile into continuum states, transfer to bound or unbound states of the target, among others \cite{j.nuclphysa.2007.05.012,j.nuclphysa.2008.01.030,i2009-10822-6,PhysRevC.84.044604,PhysRevC.81.024601,j.physletb.2010.11.007,PhysRevLett.110.142701,PhysRevC.93.044605,PhysRevLett.118.152502}. Close to or even below the Coulomb barrier, the Coulomb break-up can play an important role. In some reactions of exotic nuclei with heavy targets, Coulomb break-up can even dominate the reaction mechanisms \cite{PhysRevC.92.014604}.

Our goal is to study such Coulomb reaction effects, their interference with nuclear ones and the consequent elastic scattering flux absorption, as a function of the structural properties of different projectiles ($^{6}$He, $^{9,11}$Li, $^{9,11}$Be and $^{12}$C). With this proposal, we analyse key projectiles ranging from stable (tightly bound) to exotic, passing through stable weakly bound nuclei, reacting on different heavy targets ($^{120}$Sn, $^{197}$Au and $^{208}$Pb), at energies around the Coulomb barrier.

 The $^{12}$C nucleus is created in the stars through the so called triple $\alpha$ process triggered by the Hoyle state ($E^{*}$ = 7654.07(19) keV) \cite{KELLEY201771}. It represents a stable tightly bound ($Q_{\alpha}$ = 7366.59(4) keV) 3$\alpha$ cluster structure \cite{KELLEY201771}, which is crucial for Astrophysics, Organic Chemistry and life. 
 
 The $^9$Li exotic nucleus decays, by beta minus emission, in $^{9}$Be, with a half-life of 178.3(4) ms \cite{j.nuclphysa.2004.09.059}. The ground state is a 3/2$^{-}$. The first known excited state is 1/2$^{-}$ with excitation energy of 2691(5) keV. The next known state is a $5/2^{-}$ resonance at 4296(15) keV. The one neutron separation energy is $S_{1n}$ = 4062.22(19) keV \cite{j.nuclphysa.2004.09.059}.

Unlike $^9$Li, $^{9}$Be is a stable weakly bound nucleus with a Borromean structure composed by two $\alpha$ particles and one weakly bound neutron. The $^{9}$Be ($\alpha$+$\alpha$+$n$) nucleus has a smaller binding energy ($\varepsilon_{b}$=-1572.7(16) keV) than $^9$Li, below the $\alpha$+$\alpha$+$n$ threshold. The one neutron separation energy of $^{9}$Be ($S_{1n}$=1664.54(8) keV) \cite{j.nuclphysa.2004.09.059} is the lowest compared to other weakly bound stable nuclei and the closest to the exotic nuclei ones (table I). Thus, $^{9}$Be when colliding with a target nucleus, tends to transfer its weakly bound neutron, being this process followed by $\alpha$+$\alpha$ or $^{8}$Be, which also decays into $\alpha$+$\alpha$. 

The $^{6}$He ($\alpha$+$n$+$n$) also represents a Borromean nucleus, therefore, the three binary sub-systems, $^{4}$He-$n$ and $n$-$n$, are not bound. The $^{6}$He exotic nucleus decays, by beta minus emission, in $^{6}$Li, with a half-life of 806.7(15) ms \cite{S0375-9474(02)00597-3}. Reactions induced by $^{6}$He on different targets, at energies around the Coulomb barrier, exhibit large cross sections for $\alpha$ particles production \cite{j.nuclphysa.2007.05.012,PhysRevLett.84.5058,PhysRevC.69.044613,PhysRevC.84.044604,PhysRevC.99.054605}. It confirms a break-up picture, which is associated to the weak binding of the halo neutrons ($S_{2n}$ = 975.45(5) keV) \cite{S0375-9474(02)00597-3}, that favours the dissociation of the $^{6}$He projectile in the nuclear and Coulomb fields of the target. Thus, the two weakly bound valence neutrons modify the way in which $^{6}$He interacts, under Coulomb and/or nuclear interaction \cite{j.nuclphysa.2008.01.030,j.nuclphysa.2010.03.013,j.physletb.2010.07.060}.

The $^{11}$Be exotic nucleus decays, by beta minus emission, in $^{11}$B, with a half-life of 13.76(7) s \cite{j.nuclphysa.2012.01.010}. The low-lying spectrum of $^{11}$Be consists of two bound states: the ground state 1/2$^{+}$ and an excited state 1/2$^{-}$, at 320,04(10) keV of excitation energy \cite{j.nuclphysa.2012.01.010}. This excited state shows the largest measured $B(E1)$ distribution between bound states \cite{j.physletb.2014.03.049}. The one-neutron separation energy is $S_{1n}$ = 501.64(25) keV \cite{j.nuclphysa.2012.01.010}.

The $^{11}$Li ($^{9}$Li+$n$+$n$) represents another neutron rich exotic nucleus with Borromean structure, therefore, the three binary sub-systems, $^{9}$Li-$n$ and $n$-$n$, are not bound. The $^{11}$Li decays, by beta minus emission, in $^{11}$Be, with a half-life of 8.75(14) ms \cite{j.nuclphysa.2012.01.010}. There is no bound excited states. The two-neutron separation energy is $S_{2n}$ = 396.15(65) keV \cite{PhysRevLett.101.202501}. These two neutrons have high probability of being out of the nuclear potential range, which leads to an extended nuclear halo compared to the other $^{6,7,8,9}$Li isotopes. 
	
The one- or two-neutron structures of $^{6}$He, $^{9,11}$Li or $^{9,11}$Be can be easily polarizable in the strong electric field of a heavy target (in our case $^{120}$Sn, $^{197}$Au and $^{208}$Pb). The repulsive Coulomb long-range effect acts respectively on the cores ($^{4}$He in the case of $^{6}$He; $^{9}$Li, in the case of $^{11}$Li and $^{8,10}$Be, in the cases of $^{9,11}$Be). Notwithstanding, it does not act on the neutron(s), which implies on core repulsion while neutron(s) tend to move forward. This effect produces a distortion of the wave function that tends to reduce the Coulomb repulsion between the interacting nuclei. The reduced Coulomb repulsion decreases the elastic scattering cross section and can be described through a dynamic polarization potential (DPP), induced by dipole Coulomb excitation. This potential is then composed by two components: an attractive real part that describes the reduction of the Coulomb repulsion and an absorptive imaginary part that describes the elastic scattering cross section reduction \cite{0375-9474(94)90820-6,0375-9474(94)00765-F}. 

In \cite{PhysRevC.92.014604}, our analyses demonstrated the dominance of Coulomb dipole polarization (CDP) potential in the description of the elastic scattering angular distributions of $^{11}$Li+$^{208}$Pb. This effect is the main responsible for the observed unusual long range absorption, which is well accounted for the CDP potential. The calculated reaction cross section density (RCSD) corroborated that the absorption mostly takes place in the region of large interacting distances ($R>15$ fm).

Here, we intend to study such effect, as a function of the projectile binding energy. Thus, in Table I, we present the break-up threshold (in MeV) for different stable and exotic nuclei. 

\begin{table}[h]
%\squeezetable
\caption{\label{tab:qvalueBU} Break-up threshold of light stable and exotic nuclei.}
\begin{ruledtabular}
\begin{tabular}{   c  c c }
 system & cluster & $\varepsilon_{b}$ (MeV) \\
\hline
$^{11}$Li          & $^{9}$Li+2$n$            &  -0.396 \\ 
$^{11}$Be          & $^{10}$Be+1$n$           &  -0.502 \\
$^6$He             & $^4$He+2$n$              &  -0.975 \\
$^6$Li             & $\alpha$+$d$             &  -1.473 \\
$^9$Be             & $\alpha$+$\alpha$+1$n$   &  -1.573 \\
$^9$Be             & $^{8}$Be+1$n$            &  -1.664 \\
$^7$Li             & $\alpha$+$t$             &  -2.467 \\
$^{9}$Li           & $^{7}$Li+2$n$            &  -4.062 \\
$^{10}$B           & $^6$Li+$\alpha$          &  -4.461 \\
$^{18}$O           & $^{14}$C+$\alpha$        &  -6.228 \\
$^{16}$O           & $^{12}$C+$\alpha$        &  -7.162 \\
$^{12}$C           & $^8$Be+$\alpha$          &  -7.367 \\
%$^{12}$C^{H.s.}    & $^8$Be+$\alpha$         &  -7.653 \\
$^4$He             & $^3$He+1$n$              &  -20.6 \\

\end{tabular}
\end{ruledtabular}
\end{table} 

In section II, we present our theoretical optical model (OM) and continuum discretized coupled channel (CDCC) calculations. In section III, we compare optical potential (OP) and the trivial equivalent local potential (TELP) extracted from CDCC calculations. In addition, we compare the theoretical OM and CDCC predictions to the experimental elastic scattering angular distributions. Finally, in section IV, we present the main conclusions.  

%\maketitle          
            
%%%%%%%%%%%%%%%%%%%%%%%%%%%%%%%%%%%%%%%%%%%%%%%%%%%%%%%%%%%%%%%%%%%%%%%%%%%%%%%
\section{Theoretical approach}
\label{Data}      
%%%%%%%%%%%%%%%%%%%%%%%%%%%%%%%%%%%%%%%%%%%%%%%%%%%%%%%%%%%%%%%%%%%%%%%%%%%%%%%

%\maketitle          
            
%%%%%%%%%%%%%%%%%%%%%%%%%%%%%%%%%%%%%%%%%%%%%%%%%%%%%%%%%%%%%%%%%%%%%%%%%%%%%%%
\subsection{S\~ao Paulo potential (SPP)}
\label{Data}      
%%%%%%%%%%%%%%%%%%%%%%%%%%%%%%%%%%%%%%%%%%%%%%%%%%%%%%%%%%%%%%%%%%%%%%%%%%%%%%%

Stable nuclei reactions have been successfully described assuming the S\~ao Paulo potential (SPP) \cite{PhysRevC.66.014610}. It describes the real bare nuclear interaction and predicts, with great accuracy, experimental angular distributions of a large number of stable systems in a wide energy range, with no adjustable parameters \cite{PhysRevLett.78.3270,PhysRevLett.79.5218,PhysRevC.58.576}. 
Within this approach, the nuclear interaction, coined as  $V_{\text{SPP}}$, is written as a function of the double folding potential ($V_{\text{Fold}}$) through: 
\begin{equation} V_{\text{SPP}}(R)=V_{\text{Fold}}(R)\text{e}^{-4v^{2}/c^{2}}, \label{eq:1}
\end{equation}where $c$  is the speed of light, $v$ is the relative velocity between projectile and target, and $V_{\text{Fold}}$ is given by: 
\begin{equation}
V_{\text{Fold}}(R)=\int \int \rho_{1}(\vec{r}_{1})\rho_{2}(\vec{r}_{2})V_{0}
\delta(\vec{R}-\vec{r}_1+\vec{r}_2) \, d\vec{r}_1 \, d\vec{r}_2 . 
\label{eq:2}
\end{equation}
where $\rho_{1}$ and $\rho_{2}$ are the projectile and target matter distributions; $V_{0} \delta(\vec{r})$  is the zero-range effective interaction with $V_{0}=-456$ MeV. This $V_{0}$ value has been obtained in \cite{PhysRevC.66.014610} through a potentials systematic extracted from elastic scattering data analyses. The corresponding nucleon densities were folded with the matter distribution of the nucleon to obtain the respective matter densities \cite{PhysRevC.66.014610}. 

A fully microscopic description of the optical potential (OP), based on the Feshbach theory \cite{Fesh92,PhysRevC.55.1362,S0370-1573(96)00048-8}, is specially difficult at energies where collective as well as single particle excitation are involved in the scattering process. To face this problem, in a simple way, an extension of the SPP model to the OP imaginary part was proposed in \cite{S0375-9474(03)01158-8} and successfully applied to the elastic scattering of stable nuclei. Within this context, Eq. (1) describes the real and imaginary parts of the (nuclear) OP as follows:  
\begin{equation} V_{OP}(R)=N_RV_{\text{SPP}}(R)+iN_iV_{\text{SPP}}(R), \label{eq:3}
\end{equation}where $N_R$ and $N_i$ represent multiplication factors that determine OP strengths (real and imaginary parts) and simulate polarization (real and imaginary) effects. Therefore, Eq. (3) represents the addition of the bare and polarization potentials. The bare ($V_{\text{SPP}}(R)$) represents the ground-state expectation value of the interaction operator, which includes the average effective nucleon-nucleon force ($V_{0} \delta(\vec{r})$). The polarization part arises from nonelastic couplings. According to Feshbach theory \cite{Fesh92,PhysRevC.55.1362,S0370-1573(96)00048-8}, it is energy dependent and complex. The imaginary part arises from transitions to open non-elastic channels that absorbs flux from elastic channel. The real part arises from virtual transitions to intermediate states (inelastic excitations, nucleon transfer, among others). 

In such a scenario, assuming Eq. (3), any need of varying $N_R$ must be related to and account for contributions of the polarization potential to the OP real part, while $N_i$ accounts for the absorptive imaginary part of the polarization potential. Standard averaged values, obtained in \cite{S0375-9474(03)01158-8}, from systematical analysis of stable tightly bound nuclei are $N_R$=1.0 and $N_i$=0.78. Several recent works on stable weakly bound and exotic nuclei reactions, show how absorption processes can vary as a function of the projectile and, therefore, OP strengths \cite{PhysRevC.95.064614,PhysRevC.97.044609,PhysRevC.97.034629,PhysRevC.98.024621,PhysRevC.101.044604,PhysRevC.101.044601,PhysRevC.92.014604}. 

In \cite{PhysRevC.100.064602}, Eq. (3) has been proposed to systematically study exotic and stable, tightly and weakly bound, nuclei projectiles ($^{4,6}$He, $^{6,7}$Li, $^{9}$Be, $^{10}$B, $^{16,18}$O) reacting on the same, $^{120}$Sn, target nucleus, in a wide energy range. We verified how data are most sensitive to the strength of the absorptive imaginary part, which tends to mostly vary as a function of the projectile binding energy (table I). Therefore, table I relies on projectiles studied in \cite{PhysRevC.100.064602} and here. From \cite{PhysRevC.100.064602}, we concluded that lower the binding energy higher the absorptive imaginary strength, which tends to be meaningful even at longer interacting distances ($R>15$ fm), mainly for describing reactions of stable weakly bound and exotic nuclei projectiles. 

Thus, here, we extend such systematical study to other nuclei, besides taking explicitly into account the projectile binding energy in the optical model (OM). Once the projectile binding energy decreases, its break-up probability increases, as well as the elastic scattering flux absorption, which can also be formally taken into account within the CDCC calculations. Therefore, we dedicated next sections II.2 and II.3 to describe both approaches. 

%\maketitle          
            
%%%%%%%%%%%%%%%%%%%%%%%%%%%%%%%%%%%%%%%%%%%%%%%%%%%%%%%%%%%%%%%%%%%%%%%%%%%%%%%
\subsection{Coulomb dipole polarization (CDP) potential}
\label{Data}      
%%%%%%%%%%%%%%%%%%%%%%%%%%%%%%%%%%%%%%%%%%%%%%%%%%%%%%%%%%%%%%%%%%%%%%%%%%%%%%%

\begin{figure}[h]
\begin{minipage}{20pc}
\includegraphics[width=20pc]{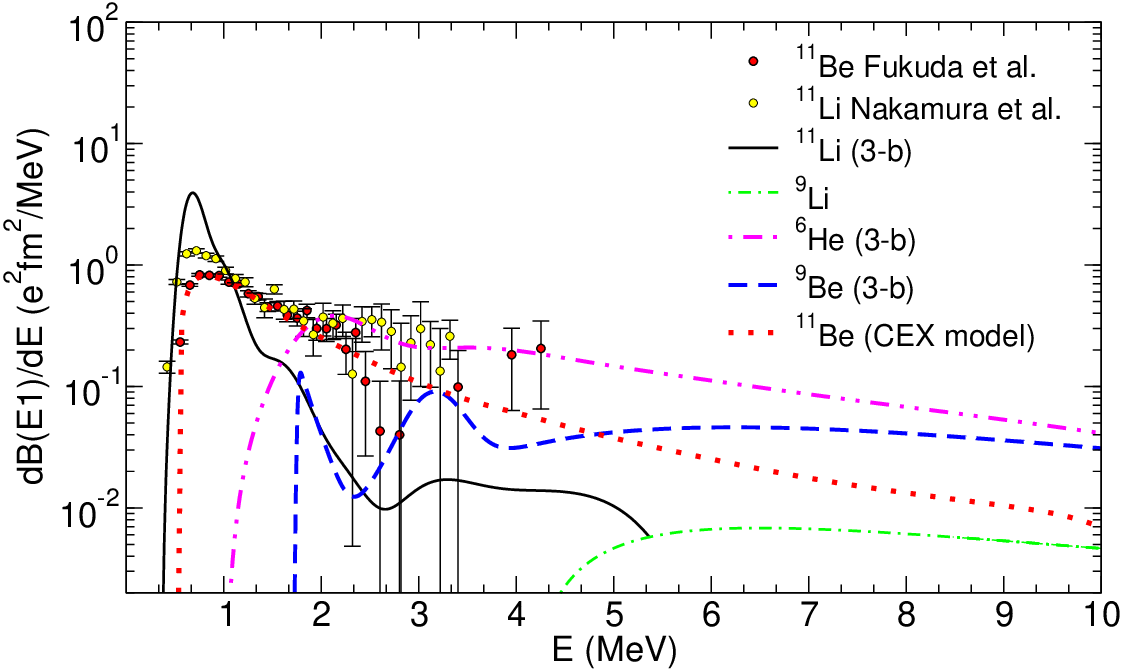}
\caption{\label{9be} $B(E1)$ experimental and theoretical distributions of $^{6}$He, $^{9,11}$Li and $^{9,11}$Be as functions of the excitation energy ($\varepsilon$) (see text for details).}
\end{minipage}\hspace{2pc}%
\end{figure}

A simple analytical formula for the CDP potential was derived in Refs.~\cite{0375-9474(94)90820-6,0375-9474(94)00765-F} and has shown to be a valuable tool to study exotic nuclei reactions in the OM context \cite{j.nuclphysa.2010.03.013}.
The CDP potential is obtained by requiring that the second order amplitude for the dipole excitation-deexcitation process and the first order amplitude, associated to the polarization potential, are equal for all classical trajectories corresponding to a given scattering energy. This leads to an analytic formula for the polarization potential according to a single excited state \cite{0375-9474(94)90820-6}. This analytic formula is generalized for the case of excitation energy to a continuum of break-up states \cite{0375-9474(94)00765-F} resulting on the following expression:

\begin{eqnarray}
U_{\text{Pol}}= - \frac{4\pi}{9} \frac{Z_{t}^{2} e^{2}}{\hbar ~ v} 
\frac{1}{(r-a_{0})^{2}r} \int_{\varepsilon_{b}}^{\infty} \text{d}\varepsilon 
\frac{\text{d}B(E1,\varepsilon)}{\text{d}\varepsilon} \times  \nonumber\\ 
\times \left[  g \left(\frac{r}{a_{0}}-1,\xi \right)+ if \left( 
\frac{r}{a_{0}}-1,\xi \right) \right]  . 
\label{eq:4}
\end{eqnarray}

Here, we can note the lineal dependence of the polarization potential with the projectile $B(E1)$ distribution and its quadratic dependence with the atomic number of the target, $Z_{\text{t}}$. Such dependence gives rise to an important contribution of the CDP, which is mainly manifested in the case of exotic nuclei reacting on heavier targets. Notwithstanding, small effects can also be observed for reactions of stable weakly bound projectiles. In Eq. (\ref{eq:4}), $\varepsilon_{b}$ is the necessary energy to break-up the projectile (table I), which is a positive value, $a_{0}$ is the half of the distance of closest approach in the head-on collision, $v$ is the velocity of the projectile  and $f$ and $g$ are analytic functions expressed as:
\begin{eqnarray}
f(z,\xi)=4\xi^{2}z^{2}\text{exp}(-\pi\xi)K_{2i\xi}''(2\xi z) \nonumber\\
g(z,\xi)=\frac{P}{\pi}\int_{-\infty}^{+\infty} \frac{f(z,\xi ')}{\xi-\xi '}\text{d}\xi '.
\label{eq:5}
\end{eqnarray}
$\xi=\frac{\varepsilon~a_{0}}{\hbar v}$ is the Coulomb adiabaticity 
parameter corresponding to the excitation energy $\varepsilon$ of the nucleus. $K''$ represents the second derivative of the Bessel functions and $P$ means the principal value of the integral. As discussed in detail in \cite{0375-9474(94)90820-6}, when the break-up energy $\varepsilon_{b}$ is large enough, the purely real adiabatic dipole potential is obtained. In the opposite limit, for small energies, 
$f (\frac{r}{a_{0}}-1,\xi)\rightarrow 1$ and $g (\frac{r}{a_{0}}-1,\xi)
\rightarrow 0$, and the polarization potential becomes purely imaginary, 
depending on $r$ as $\frac{1}{(r-a_{0})^{2}r}$.

Figure 1 shows $B(E1)$ experimental and theoretical distributions of $^{6}$He, $^{9,11}$Li and $^{9,11}$Be as functions of the excitation energy ($\varepsilon$). The $^{11}$Li and $^{11}$Be $B(E1)$ experimental values were reported in \cite{PhysRevLett.96.252502} and \cite{PhysRevC.70.054606}. The solid line corresponds to the three-body model calculation, for $^{11}$Li, reported in \cite{PhysRevLett.110.142701}. The dotted line corresponds to the core excitation model calculation, for $^{11}$Be $B(E1)$ distribution, reported in \cite{PhysRevLett.118.152502}. The dashed line corresponds to the three-body model calculation for the $^{9}$Be $B(E1)$ distribution reported in \cite{PhysRevC.90.044304}. The dashed-double-dotted line corresponds to the three-body model calculation, for $^{6}$He $B(E1)$ distribution, reported in \cite{PhysRevC.92.014604}. The dashed-dotted line corresponds to a cluster model ($^{7}$Li+2n) \cite{PhysRevLett.49.1482,0375-9474(91)90442-9}, considered for calculating the $^{9}$Li $B(E1)$ distribution.

%\maketitle          
            
%%%%%%%%%%%%%%%%%%%%%%%%%%%%%%%%%%%%%%%%%%%%%%%%%%%%%%%%%%%%%%%%%%%%%%%%%%%%%%%
\subsection{Continuum Discretized Coupled Channel (CDCC) calculations}
\label{Data}      
%%%%%%%%%%%%%%%%%%%%%%%%%%%%%%%%%%%%%%%%%%%%%%%%%%%%%%%%%%%%%%%%%%%%%%%%%%%%%%%
The coupling to break-up channels in reactions involving weakly bound nuclei can be formally described within the CDCC formalism~\cite{10.1143/PTPS.89.32,0370-1573(87)90094-9}, in which the total scattering wave function is expanded in internal states of the projectile in a given few-body model. This is typically referred to as three-body CDCC (for two-body projectiles) or four-body CDCC (for three-body projectiles). %In general, the projectile $+$ target wave function can be written as
%\begin{equation}
%\begin{split}
%\Psi_{JM}(\boldsymbol{R},\boldsymbol{\xi})  \equiv & \sum_{nj\mu LM_L} \phi_{nj\mu}(\boldsymbol{\xi})\langle LM_L j\mu|JM\rangle  \\
                                           %                    & \times i^L Y_{LM_L}(\widehat{R})\frac{1}{R}f_{Lnj}^J(R)
%\end{split}
%\end{equation}
The corresponding coupling form factors are generated following a multipole expansion (in $Q$ multipoles) of the projectile-target interaction, using as input suitable fragment-target optical potentials. 

In the present work, for $^{11}$Li, $^6$He and $^9$Be nuclei reactions we perform four-body CDCC calculations using the three-body structure models ($^9\text{Li}+n+n$, $\alpha+n+n$, and $\alpha+\alpha+n$) and fragment-target optical potentials presented in Refs.~\cite{PhysRevLett.109.262701,PhysRevC.81.044605,PhysRevC.92.054611}. In this sense, the calculations here presented have no free-parameters. The states of $^{11}$Li and $^{6}$He are computed using a binning procedure (Ref.~\cite{PhysRevC.80.051601}) from the true three-body continuum states, whereas $^{9}$Be states are generated using a pseudostate approach, the analytical Transformed Harmonic Oscillator method presented in Ref.~\cite{PhysRevC.88.014327}. With these ingredients, the CDCC problem solved up to convergence in the number of partial waves and excitation energy above the respective break-up thresholds, including continuum couplings to all multipole orders. This allows us to dissect the behavior of the elastic scattering cross section in terms of different contributions, from no-continuum calculations to the effect of dipole and higher order contributions, as discussed in the next section. 

In the reaction $^{11}$Be+$^{197}$Au, we consider $^{11}$Be as a core+valence ($^{10}$Be+n) two-body system where we allow the $^{10}$Be core to be excited to its first excited state ~\cite{PhysRevC.85.054618}. To include this structure properly in the reaction mechanism, we perform a CDCC calculation including core excitations (XCDCC)~\cite{PhysRevC.89.064609,PhysRevC.74.014606}. We use the same fragment-target optical potentials and also the potential for the structure of $^{11}$Be as those presented in~\cite{PhysRevLett.118.152502}. Following~\cite{PhysRevLett.118.152502}, we also apply a binning procedure for the discretization of the continuum. 

In all cases, we compute the trivial equivalent local potentials (TELP), which have the meaning of simple optical potentials (real and imaginary parts) leading to the same elastic scattering cross section obtained by the CDCC. The TELP is calculated following the prescription proposed in~\cite{THOMPSON198984}, which involves two steps. First, for each total angular momentum, a trivially-equivalent local polarization potential is calculated from the source term of the elastic channel equation. Then, an approximate TELP is constructed by averaging these L-dependent polarization potentials, using as weights the transfer/breakup cross section for each angular momentum. The TELP obtained by this procedure can be regarded as a L-independent local approximation of a complicated coupled-channels system. If the TELP extracted from the coupled-channels calculation is a good representation of the overall effect of the couplings, the solution of the single-channel Schrödinger equation with the effective potential $U_{\text{eff}}$=$U_{\text{bare}}$+$U_{\text{TELP}}$ should reproduce the elastic scattering similar to the one obtained with the full coupled-channels calculation. The bare potential, $U_{\text{bare}}$, is just the sum of the fragment–target interactions convoluted with the ground state density of the projectile nucleus. These potentials, at large distances, can be compared with the CDP part of the OM approach in order to corroborate the observed trends. Note that, within the CDCC formalism, the influence of the projectile binding energy on the reaction dynamics is implicitly contained in the corresponding coupling potentials. If the ground state is closer to the threshold, its wave function explores larger distances, thus leading to stronger couplings with the low-lying continuum states. This, in first order, produces larger break-up cross sections, therefore, we expect a larger imaginary part of the TELP for less bound systems. 

%%%%%%%%%%%%%%%%%%%%%%%%%%%%%%%%%%%%%%%%%%%%%%%%%%%%%%%%%%%%%%%%%%%%%%%%%%%%%%%
\section{Data Analysis}
\label{Data}      
%%%%%%%%%%%%%%%%%%%%%%%%%%%%%%%%%%%%%%%%%%%%%%%%%%%%%%%%%%%%%%%%%%%%%%%%%%%%%%%

The OM approach assumed here is written in terms of the SPP (Eq. 3) and CDP (Eq. 4) models, through:
\begin{equation}
U_{\text{OP}}(R)=N_{R}V_{\text{SPP}}(R)+iN_{i}V_{\text{SPP}}(R)+V_{\text{Pol}}(R)+
iW_{\text{Pol}}(R).\label{eq:6}
\end{equation}
$V_{\text{SPP}}$ is given by Eq. (1); $V_{\text{Pol}}$ and $W_{\text{Pol}}$ represent the real and imaginary terms of $U_{\text{Pol}}$, according to Eq.~(\ref{eq:4}); $N_{R}$ = 1.00 and $N_{i}$ = 0.78 are reference values obtained in \cite{S0375-9474(03)01158-8} and applied in the current work. 

\begin{figure}[h]
\begin{minipage}{20pc}
\includegraphics[width=20pc]{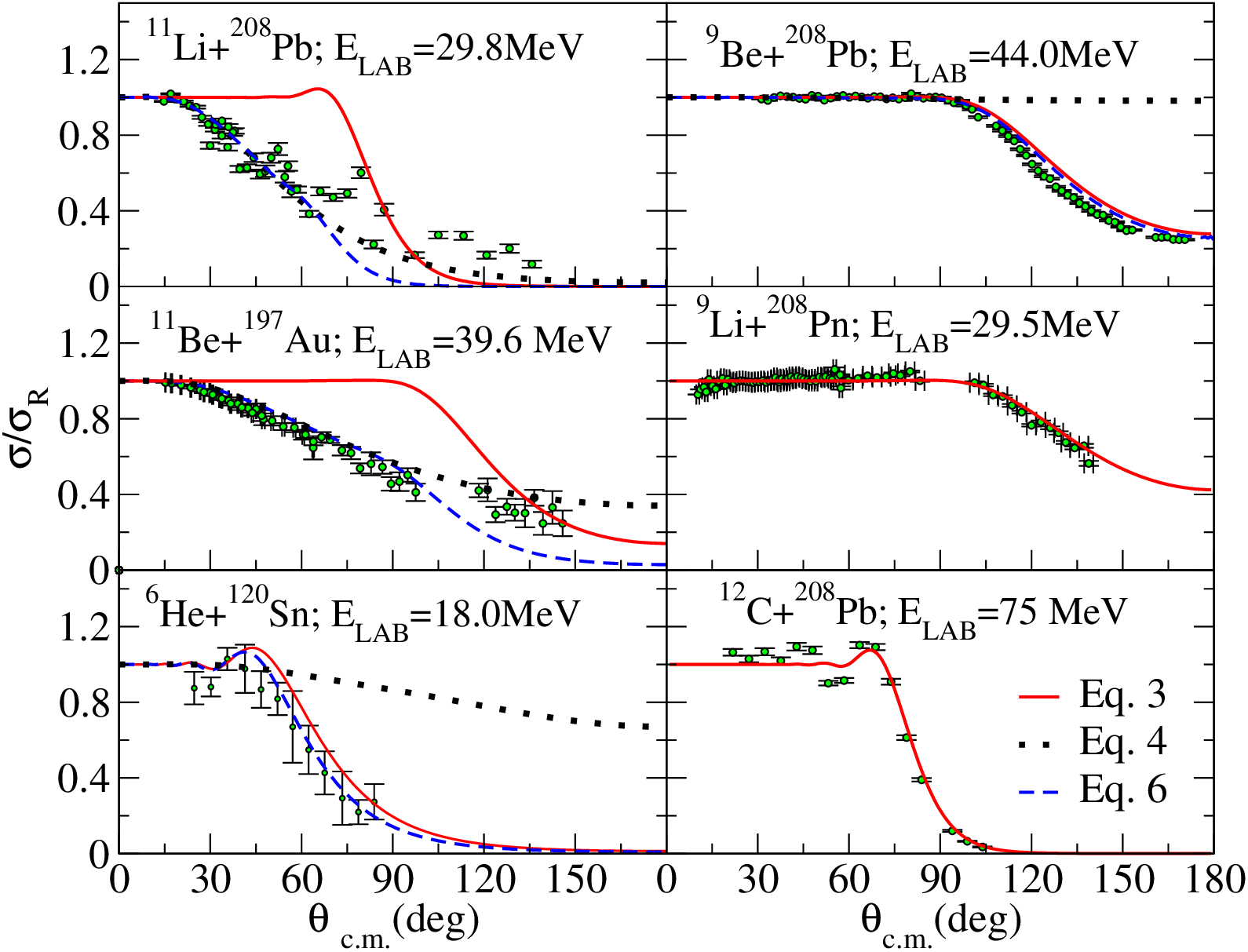}
\caption{\label{9be} Elastic scattering angular distributions of $^{12}$C, $^{9}$Be and $^{9,11}$Li+$^{208}$Pb, $^{11}$Be+$^{197}$Au and $^{6}$He+$^{120}$Sn. Experimental data are extracted from \cite{S0375-9474(03)01158-8,Yu075108,PhysRevLett.109.262701,PhysRevLett.118.152502,PhysRevC.82.044606}. Curves represent OM calculations based on Eq. 3 (solid line; only SPP), Eq. 4 (dotted line; only CDP), and Eq. 6 (dashed line; SPP+CDP).}
\end{minipage}\hspace{2pc}%
\end{figure}

\begin{figure}[h]
\begin{minipage}{20pc}
\includegraphics[width=20pc]{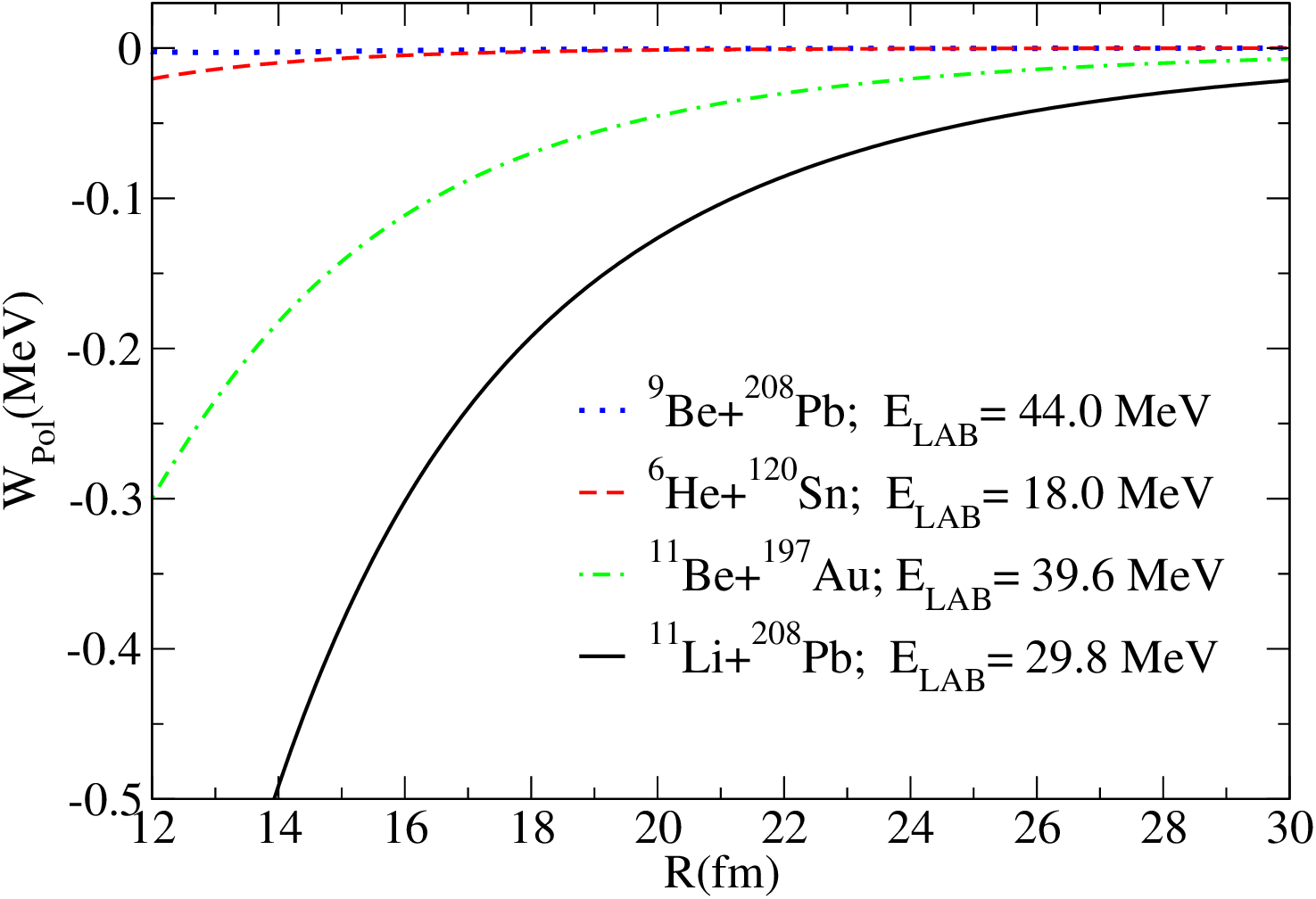}
\caption{\label{LA_grafica} The strength of the CDP imaginary potential ($W_ {Pol}$), given by the imaginary part of Eq. (4), as a function of the interacting distance ($R$).}
\end{minipage}\hspace{2pc}%
\end{figure}

\begin{figure}[h]
\begin{minipage}{20pc}
\includegraphics[width=20pc]{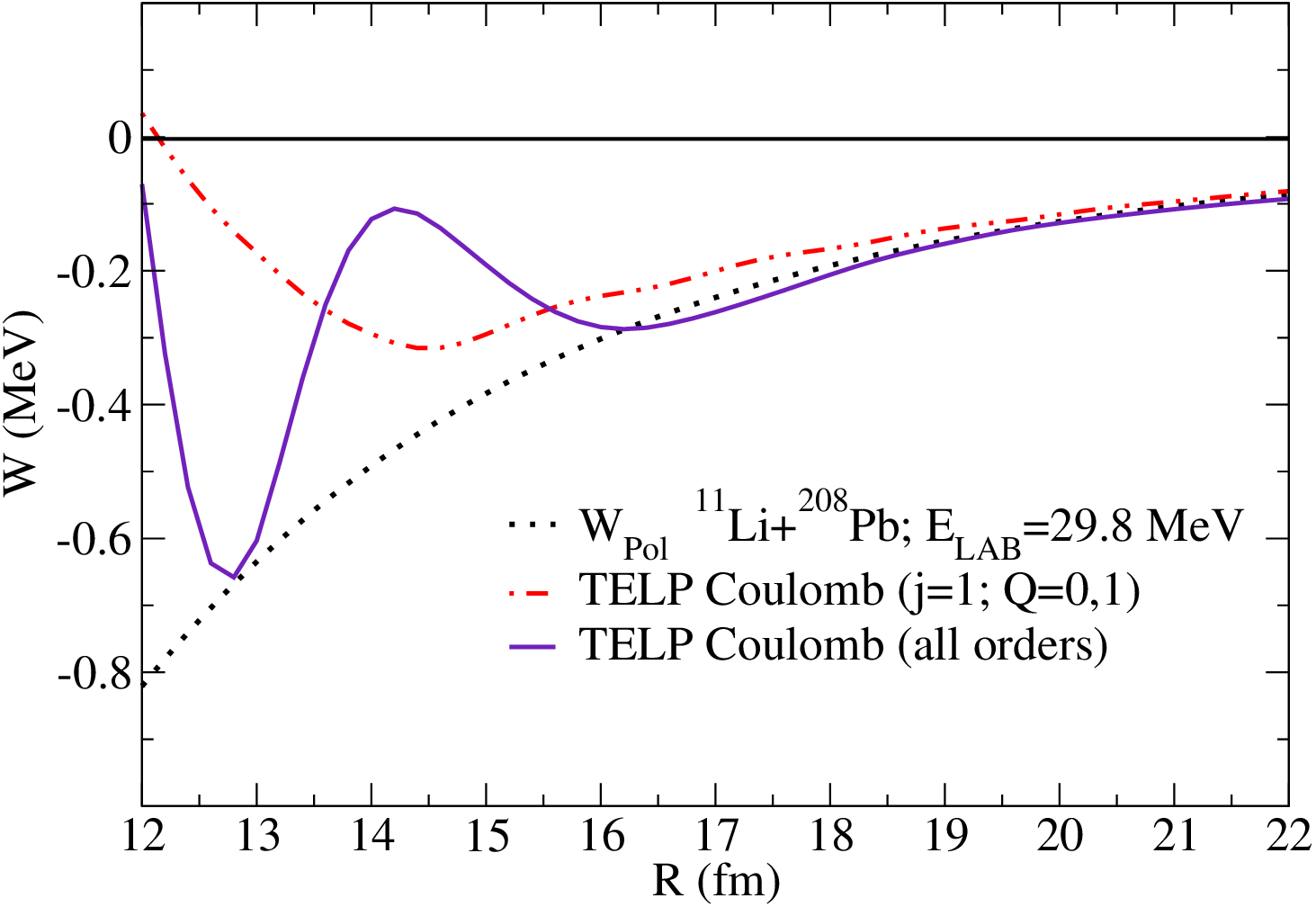}
\caption{\label{LA_grafica} The strength of the CDP imaginary potential, obtained from different approaches. Dotted line represents the term $W_{\text{Pol}}(R)$ of Eq. (4), calculated for the $^{11}$Li+$^{208}$Pb system, at $E_{LAB}=29.8$ MeV. Dashed-dotted line represents the imaginary TELP for Coulomb monopolar and dipolar contributions ($j=1; Q=0,1$). Solid line represents the imaginary TELP for all Coulomb contributions.}
\end{minipage}\hspace{2pc}%
\end{figure}

\begin{figure}[h]
\begin{minipage}{20pc}
\includegraphics[width=20pc]{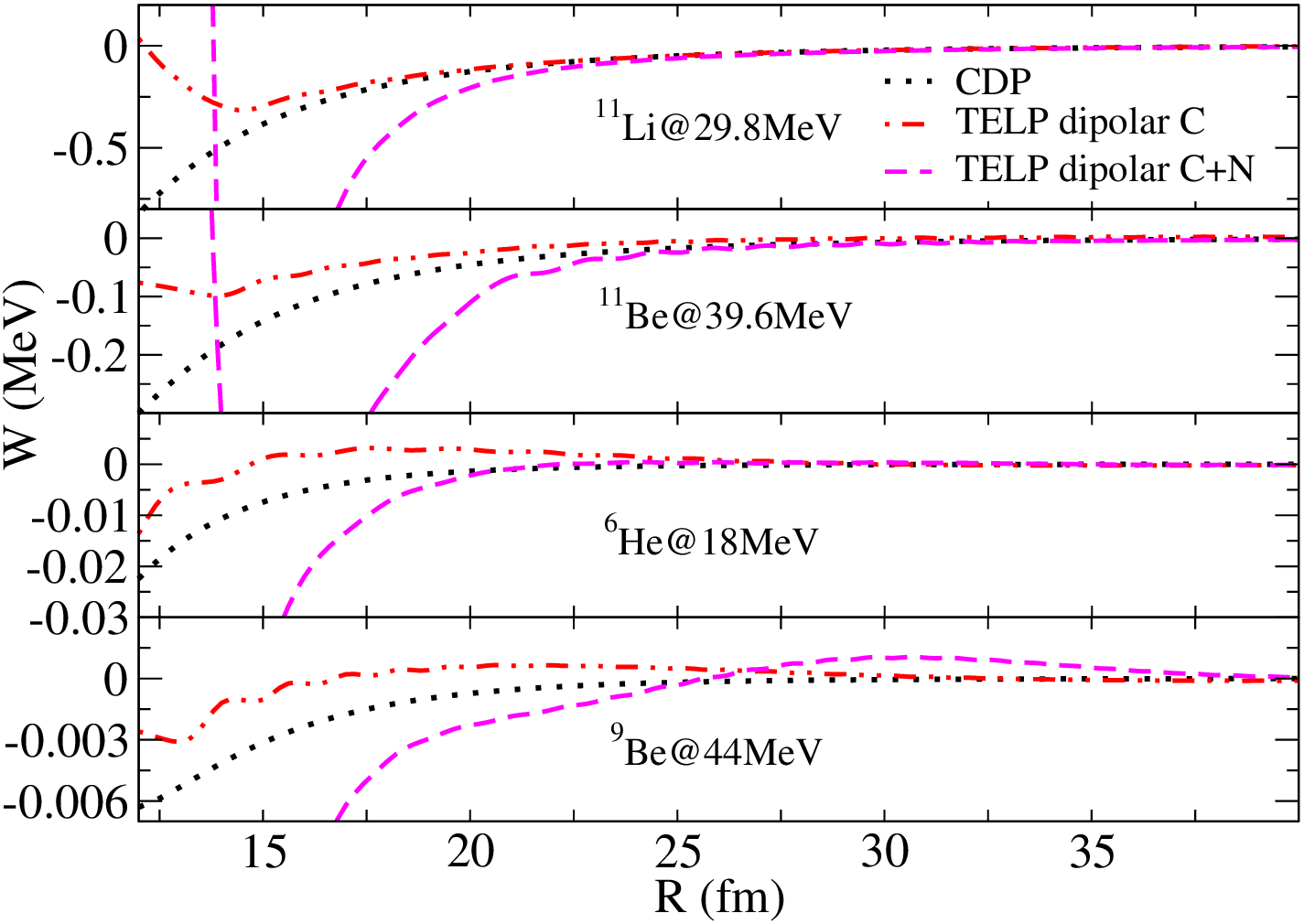}
\caption{\label{LA_grafica} The strength of the CDP imaginary potential ($W_ {Pol}$), in Eq. (4), as a function of the interacting distance ($R$) (dotted line), compared to the TELP, considering only Coulomb dipolar effect (dashed-dotted line) and both Coulomb and nuclear dipolar effects (dashed line).}
\end{minipage}\hspace{2pc}%
\end{figure}

\begin{figure}[h]
\begin{minipage}{20pc}
\includegraphics[width=20pc]{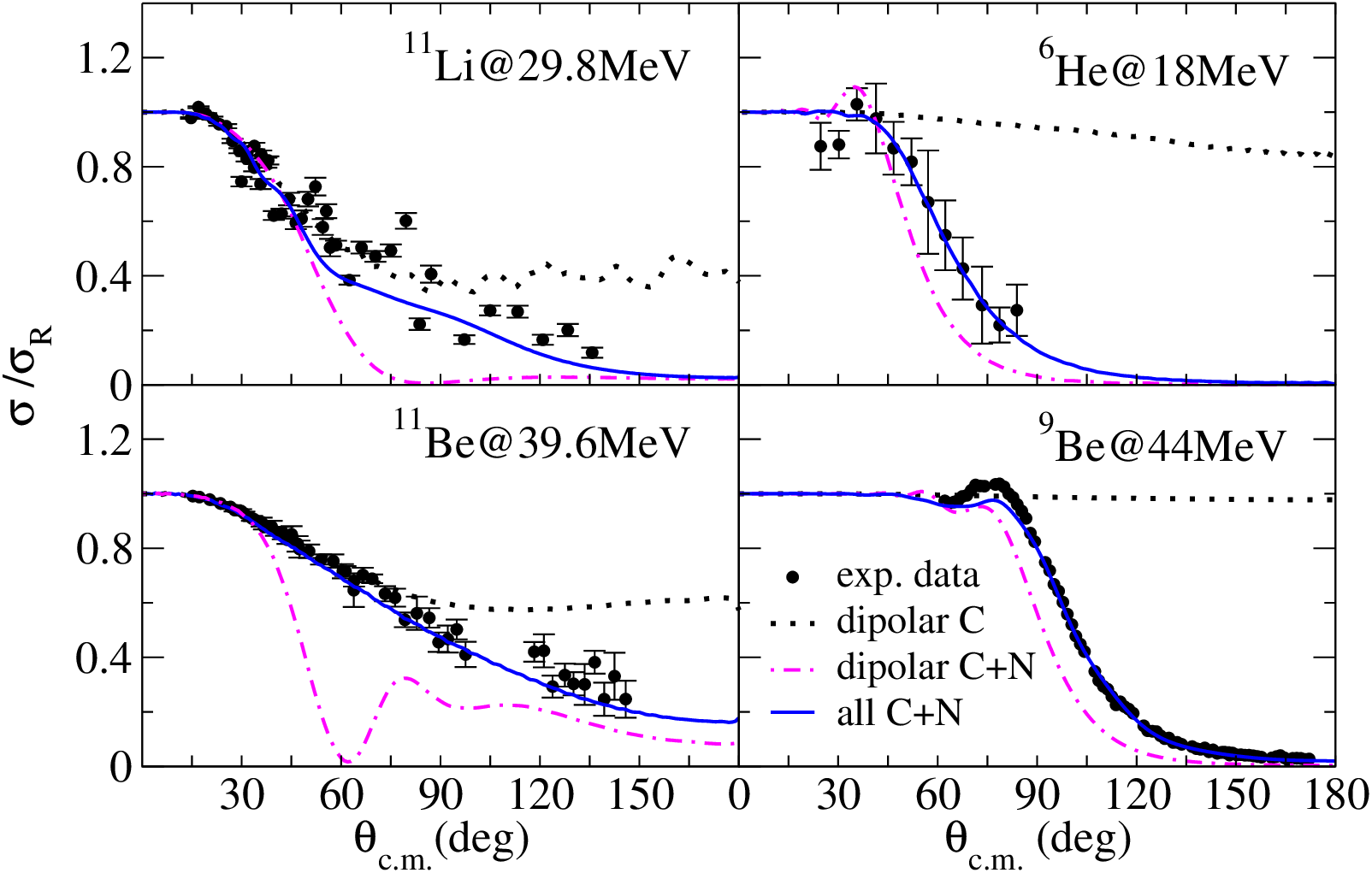}
\caption{\label{9be} Elastic scattering angular distributions of $^{9}$Be and $^{11}$Li+$^{208}$Pb, $^{11}$Be+$^{197}$Au and $^{6}$He+$^{120}$Sn. Experimental data are extracted from \cite{PhysRevLett.109.262701,PhysRevLett.118.152502,PhysRevC.82.044606,Yu075108}. Curves represent CDCC calculations considering different contributions. In particular, for $^{11}$Be, solid line was published in \cite{PhysRevLett.118.152502}.}
\end{minipage}\hspace{2pc}%
\end{figure}

Therefore, in Eq. (6), the OP real part ($V(R)$) is represented by the sum:
\begin{equation}
V(R)=V_{\text{SPP}}(R)+V_{\text{Pol}}(R),\label{eq:7}
\end{equation}
while, the OP imaginary part ($iW(R)$) is represented by the sum:
\begin{equation}
iW(R)=i(0.78)V_{\text{SPP}}(R)+iW_{\text{Pol}}(R).\label{eq:8}
\end{equation}

In Fig. 2, this OM approach is applied to describe elastic scattering angular distributions of $^{12}$C, $^{9}$Be and $^{9,11}$Li+$^{208}$Pb, $^{11}$Be+$^{197}$Au and $^{6}$He+$^{120}$Sn. For $^{12}$C,$^{9}$Li+$^{208}$Pb systems, OM calculations, considering Eq. (6) and Eq. (3), are exactly the same (solid lines). Dotted lines represent only the absorption effect due to the CDP (given by Eq. 4) in the calculations. Such effect is negligible, for $^{12}$C,$^{9}$Li+$^{208}$Pb systems, however, it tends to increase and dominates absorption processes as the projectile binding energy decreases (see table I). Thus, from Fig. 2 we can observe (dotted lines) how the CDP potential becomes important for $^{9}$Be+$^{208}$Pb, increases for $^{6}$He+$^{120}$Sn, and dominates for $^{11}$Be+$^{197}$Au and $^{11}$Li+$^{208}$Pb systems. Moreover, there is a remarkable agreement between data and theoretical calculations performed with Eq. (6). It is worth to mention that all calculations were performed without any free-parameter. 

This absorption effect can be better appreciated in Fig. 3, where we plot the strength of the CDP imaginary potential ($W_ {Pol}$), as a function of the interacting distance ($R$), for different systems. As the projectile binding energy decreases, this strength increases and becomes stronger at longer distances. The most pronounced effect is observed for the $^{11}$Li+$^{208}$Pb. 

Thus, in Fig. 4, we compare the strength of the CDP imaginary potential ($W_ {Pol}$) (dotted line) with the imaginary Coulomb ($j=1; Q=0,1$) TELP (dashed-dotted line) and the imaginary Coulomb (all contributions) TELP (solid line), for the $^{11}$Li+$^{208}$Pb. Calculations converge to similar strengths, at distances greater than $16$ fm, and, calculations fully converge for distances greater than $20$ fm. Therefore, Coulomb dipolar effect shows to dominate the absorption mechanism in $^{11}$Li reactions.  

In Fig. 5, for $^{11}$Li, $^{11,9}$Be and $^{6}$He reacting on heavy targets, we compare the strength of the CDP imaginary potential ($W_ {Pol}$) (dotted line) with TELP, extracted from CDCC calculations, considering only Coulomb dipolar effect (dashed-dotted line) and both Coulomb and nuclear dipolar effects (dashed line). The former involves the ground state and the continuum states of the projectile compatible with dipole excitations, including $Q=0,1$ terms of the multipole expasion of the Coulomb interaction. The latter is the same but adding also the nuclear term. It is worth to mention that the conclusions extracted from the TELP have to be analyzed with caution, since, as it was previously commented, this potential is just a local L-independent approximation of a very complicated non-local and L-dependent object. Notwithstanding, as it was shown in \cite{j.nuclphysa.2010.03.013}, nuclear couplings are also responsible for the strong repulsive part of the polarization potential. Although they are of shorter range than the Coulomb polarization potential, it was verified that both the real and imaginary components of the nuclear polarization potential extend to distances well beyond the strong absorption radius. Therefore, besides long range Coulomb couplings, light exotic nuclei reactions, mainly on heavy targets, are characterized by long range nuclear couplings. These features are also consistent with the findings of Mackintosh and Keeley \cite{PhysRevC.79.014611} and Rusek \cite{RUSEK}. Furthermore, in \cite{PhysRevC.79.014611}, it has been pointed out that an emissive imaginary part can appear as a consequence of representing a strongly nonlocal object, namely, the dynamic polarization potential arising from the coupled-channels, by a simple local potential. This effect, nevertheless, does not lead to unitary breaking. In Fig. 5 curves are quite different at $R=16$ fm and just converge (to zero) at $R>25$ fm. Such differences are crucial to cross sections determination, as it can be verified in Fig. 6.

Finally, in Fig. 6, we perform CDCC calculations to describe elastic scattering angular distributions of $^{9}$Be and $^{11}$Li+$^{208}$Pb, $^{11}$Be+$^{197}$Au and $^{6}$He+$^{120}$Sn. These calculations correspond to the same prescriptions presented in Fig. 5 for TELP. For each system, we can observe the Coulomb dipolar effect (dotted line). As in Fig. 2, such Coulomb dipolar effect is much more pronounced for $^{11}$Li and $^{11}$Be. For the other two projectiles, $^6$He and $^9$Be, nuclear dipolar effects (dashed-dotted line) play a major role. Notwithstanding, for all systems, the combination of Coulomb and nuclear effects, considering contributions of all orders (solid line), is necessary to describe the overall data trend with better accuracy. These calculations involve the ground state and all continuum states of the projectile, including all Q terms of the interaction, up to convergence.  Specific details of these calculations can be found in the Refs. \cite{PhysRevLett.109.262701} (11Li), \cite{PhysRevLett.118.152502} (11Be), \cite{PhysRevC.81.044605} (6He) and \cite{PhysRevC.81.024601} (9Be).  

%\textbf{For $^{11}$Be\dots}
%The full calculation for the case of $^{11}$Be+$^{197}$Au is the same as~\cite{PhysRevLett.118.152502}. In this particular case the model space for $^{11}$Be to obtain a converged result is large and even includes non-negligible closed channels. This requires orthogonalization techniques that prevents us from obtaining the corresponding TELP~\cite{moro}.

\vspace{1.0cm}

%%%%%%%%%%%%%%%%%%%%%%%%%%%%%%%%%%%%%%%%%%%%%%%%%%%%%%%%%%%%%%%%%%%%%%%%%%%%%%%
\section{Conclusions}
\label{Data}      
%%%%%%%%%%%%%%%%%%%%%%%%%%%%%%%%%%%%%%%%%%%%%%%%%%%%%%%%%%%%%%%%%%%%%%%%%%%%%%%

This manuscript reports on systematical optical model (OM) and continuum discretized coupled channel (CDCC) calculations applied to describe the elastic scattering angular distributions of $^{6}$He, $^{9,11}$Li, $^{9,11}$Be and $^{12}$C projectiles on different heavy targets. The OM analysis was carried out within the context of the optical potential (OP) given by Eq. (6), which is based on the nuclear double folding S\~ao Paulo potential (SPP) (Eq. (3)) and the Coulomb dipole polarization (CDP) potential (Eq. (4)). Trivial equivalent local potential (TELP), derived from CDCC calculations, corroborate the results obtained with the OM approach. In addition, CDCC calculations are able to distinguish the importance of Coulomb and/or nuclear effects, of different orders, for
each specific nuclear reaction. Furthermore, CDCC calculations (Fig. 6) show the importance of considering Coulomb and nuclear effects, of all orders, with the aim of describing the overall data trend. All calculations presented here were performed without any free-parameter. Thus, the OM and CDCC predictions establish a common basis for exotic and stable nuclei reactions, accounting for important differences in their reaction mechanisms, which depends on their structural properties. 

\begin{acknowledgments} 
This work was supported by the Ministry of Science, Innovation and Universities of Spain, through the project PGC2018-096994-B-C21. This work was also partially supported by the Spanish Ministry
of Economy and Competitiveness, the European Regional Development Fund (FEDER), under Project $\rm N^o$  FIS2017-88410-P and by the European Union's Horizon 2020 research and innovation program, under Grant Agreement $\rm N^o$ 654002. J.C.~acknowledges support by SID funds 2019 (Investimento Strategico di Dipartimento, Universittà degli Studi di Padova, Italy) under project No.~CASA\_SID19\_01.
\end{acknowledgments}     

\vspace{1.0cm}
\newpage

\bibliographystyle{apsrev4-1}
\bibliography{biblio}

%merlin.mbs apsrev4-1.bst 2010-07-25 4.21a (PWD, AO, DPC) hacked
%Control: key (0)
%Control: author (72) initials jnrlst
%Control: editor formatted (1) identically to author
%Control: production of article title (-1) disabled
%Control: page (0) single
%Control: year (1) truncated
%Control: production of eprint (0) enabled
\begin{thebibliography}{63}%
\makeatletter
\providecommand \@ifxundefined [1]{%
 \@ifx{#1\undefined}
}%
\providecommand \@ifnum [1]{%
 \ifnum #1\expandafter \@firstoftwo
 \else \expandafter \@secondoftwo
 \fi
}%
\providecommand \@ifx [1]{%
 \ifx #1\expandafter \@firstoftwo
 \else \expandafter \@secondoftwo
 \fi
}%
\providecommand \natexlab [1]{#1}%
\providecommand \enquote  [1]{``#1''}%
\providecommand \bibnamefont  [1]{#1}%
\providecommand \bibfnamefont [1]{#1}%
\providecommand \citenamefont [1]{#1}%
\providecommand \href@noop [0]{\@secondoftwo}%
\providecommand \href [0]{\begingroup \@sanitize@url \@href}%
\providecommand \@href[1]{\@@startlink{#1}\@@href}%
\providecommand \@@href[1]{\endgroup#1\@@endlink}%
\providecommand \@sanitize@url [0]{\catcode `\\12\catcode `\$12\catcode
  `\&12\catcode `\#12\catcode `\^12\catcode `\_12\catcode `\%12\relax}%
\providecommand \@@startlink[1]{}%
\providecommand \@@endlink[0]{}%
\providecommand \url  [0]{\begingroup\@sanitize@url \@url }%
\providecommand \@url [1]{\endgroup\@href {#1}{\urlprefix }}%
\providecommand \urlprefix  [0]{URL }%
\providecommand \Eprint [0]{\href }%
\providecommand \doibase [0]{http://dx.doi.org/}%
\providecommand \selectlanguage [0]{\@gobble}%
\providecommand \bibinfo  [0]{\@secondoftwo}%
\providecommand \bibfield  [0]{\@secondoftwo}%
\providecommand \translation [1]{[#1]}%
\providecommand \BibitemOpen [0]{}%
\providecommand \bibitemStop [0]{}%
\providecommand \bibitemNoStop [0]{.\EOS\space}%
\providecommand \EOS [0]{\spacefactor3000\relax}%
\providecommand \BibitemShut  [1]{\csname bibitem#1\endcsname}%
\let\auto@bib@innerbib\@empty
%</preamble>
\bibitem [{\citenamefont {Burbidge}\ \emph {et~al.}(1957)\citenamefont
  {Burbidge}, \citenamefont {Burbidge}, \citenamefont {Fowler},\ and\
  \citenamefont {Hoyle}}]{RevModPhys.29.547}%
  \BibitemOpen
  \bibfield  {author} {\bibinfo {author} {\bibfnamefont {E.~M.}\ \bibnamefont
  {Burbidge}}, \bibinfo {author} {\bibfnamefont {G.~R.}\ \bibnamefont
  {Burbidge}}, \bibinfo {author} {\bibfnamefont {W.~A.}\ \bibnamefont
  {Fowler}}, \ and\ \bibinfo {author} {\bibfnamefont {F.}~\bibnamefont
  {Hoyle}},\ }\href {\doibase 10.1103/RevModPhys.29.547} {\bibfield  {journal}
  {\bibinfo  {journal} {Rev. Mod. Phys.}\ }\textbf {\bibinfo {volume} {29}},\
  \bibinfo {pages} {547} (\bibinfo {year} {1957})}\BibitemShut {NoStop}%
\bibitem [{\citenamefont {Demetriou}\ \emph {et~al.}(2002)\citenamefont
  {Demetriou}, \citenamefont {Grama},\ and\ \citenamefont
  {Goriely}}]{S0375-9474(02)00756-X}%
  \BibitemOpen
  \bibfield  {author} {\bibinfo {author} {\bibfnamefont {P.}~\bibnamefont
  {Demetriou}}, \bibinfo {author} {\bibfnamefont {C.}~\bibnamefont {Grama}}, \
  and\ \bibinfo {author} {\bibfnamefont {S.}~\bibnamefont {Goriely}},\ }\href
  {\doibase 10.1016/S0375-9474(02)00756-X} {\bibfield  {journal} {\bibinfo
  {journal} {Nucl. Phys. A}\ }\textbf {\bibinfo {volume} {707}},\ \bibinfo
  {pages} {253} (\bibinfo {year} {2002})}\BibitemShut {NoStop}%
\bibitem [{\citenamefont {Bauge}\ and\ \citenamefont
  {Dupuis}(2004)}]{1.1737132}%
  \BibitemOpen
  \bibfield  {author} {\bibinfo {author} {\bibfnamefont {E.}~\bibnamefont
  {Bauge}}\ and\ \bibinfo {author} {\bibfnamefont {M.}~\bibnamefont {Dupuis}},\
  }\href {\doibase 10.1063/1.1737132} {\bibfield  {journal} {\bibinfo
  {journal} {AIP Conference Proceedings}\ }\textbf {\bibinfo {volume} {704}},\
  \bibinfo {pages} {385} (\bibinfo {year} {2004})}\BibitemShut {NoStop}%
\bibitem [{\citenamefont {Gasques}\ \emph {et~al.}(2007)\citenamefont
  {Gasques}, \citenamefont {Afanasjev}, \citenamefont {Beard}, \citenamefont
  {Lubian}, \citenamefont {Neff}, \citenamefont {Wiescher},\ and\ \citenamefont
  {Yakovlev}}]{PhysRevC.76.045802}%
  \BibitemOpen
  \bibfield  {author} {\bibinfo {author} {\bibfnamefont {L.~R.}\ \bibnamefont
  {Gasques}}, \bibinfo {author} {\bibfnamefont {A.~V.}\ \bibnamefont
  {Afanasjev}}, \bibinfo {author} {\bibfnamefont {M.}~\bibnamefont {Beard}},
  \bibinfo {author} {\bibfnamefont {J.}~\bibnamefont {Lubian}}, \bibinfo
  {author} {\bibfnamefont {T.}~\bibnamefont {Neff}}, \bibinfo {author}
  {\bibfnamefont {M.}~\bibnamefont {Wiescher}}, \ and\ \bibinfo {author}
  {\bibfnamefont {D.~G.}\ \bibnamefont {Yakovlev}},\ }\href {\doibase
  10.1103/PhysRevC.76.045802} {\bibfield  {journal} {\bibinfo  {journal} {Phys.
  Rev. C}\ }\textbf {\bibinfo {volume} {76}},\ \bibinfo {pages} {045802}
  (\bibinfo {year} {2007})}\BibitemShut {NoStop}%
\bibitem [{\citenamefont {Alvarez}\ \emph {et~al.}(2001)\citenamefont
  {Alvarez}, \citenamefont {Rossi}, \citenamefont {Silva}, \citenamefont
  {Gasques}, \citenamefont {Chamon}, \citenamefont {Pereira}, \citenamefont
  {Rao}, \citenamefont {Carlson}, \citenamefont {De~Conti}, \citenamefont
  {Anjos}, \citenamefont {Gomes}, \citenamefont {Lubian}, \citenamefont
  {Kailas}, \citenamefont {Chatterjee},\ and\ \citenamefont
  {Singh}}]{PhysRevC.65.014602}%
  \BibitemOpen
  \bibfield  {author} {\bibinfo {author} {\bibfnamefont {M.~A.~G.}\
  \bibnamefont {Alvarez}}, \bibinfo {author} {\bibfnamefont {E.~S.}\
  \bibnamefont {Rossi}}, \bibinfo {author} {\bibfnamefont {C.~P.}\ \bibnamefont
  {Silva}}, \bibinfo {author} {\bibfnamefont {L.~R.}\ \bibnamefont {Gasques}},
  \bibinfo {author} {\bibfnamefont {L.~C.}\ \bibnamefont {Chamon}}, \bibinfo
  {author} {\bibfnamefont {D.}~\bibnamefont {Pereira}}, \bibinfo {author}
  {\bibfnamefont {M.~N.}\ \bibnamefont {Rao}}, \bibinfo {author} {\bibfnamefont
  {B.~V.}\ \bibnamefont {Carlson}}, \bibinfo {author} {\bibfnamefont
  {C.}~\bibnamefont {De~Conti}}, \bibinfo {author} {\bibfnamefont {R.~M.}\
  \bibnamefont {Anjos}}, \bibinfo {author} {\bibfnamefont {P.~R.~S.}\
  \bibnamefont {Gomes}}, \bibinfo {author} {\bibfnamefont {J.}~\bibnamefont
  {Lubian}}, \bibinfo {author} {\bibfnamefont {S.}~\bibnamefont {Kailas}},
  \bibinfo {author} {\bibfnamefont {A.}~\bibnamefont {Chatterjee}}, \ and\
  \bibinfo {author} {\bibfnamefont {P.}~\bibnamefont {Singh}},\ }\href
  {\doibase 10.1103/PhysRevC.65.014602} {\bibfield  {journal} {\bibinfo
  {journal} {Phys. Rev. C}\ }\textbf {\bibinfo {volume} {65}},\ \bibinfo
  {pages} {014602} (\bibinfo {year} {2001})}\BibitemShut {NoStop}%
\bibitem [{\citenamefont {Escrig}\ \emph {et~al.}(2007)\citenamefont {Escrig},
  \citenamefont {S\'{a}nchez-Ben\'{i}tez}, \citenamefont {Moro}, \citenamefont
  {Alvarez}, \citenamefont {Andr\'{e}s}, \citenamefont {Angulo}, \citenamefont
  {Borge}, \citenamefont {Cabrera}, \citenamefont {Cherubini}, \citenamefont
  {Demaret}, \citenamefont {Espino}, \citenamefont {Figuera}, \citenamefont
  {Freer}, \citenamefont {Garc\'{i}a-Ramos}, \citenamefont {G\'{o}mez-Camacho},
  \citenamefont {Gulino}, \citenamefont {Kakuee}, \citenamefont {Martel},
  \citenamefont {Metelko}, \citenamefont {P\'{e}rez-Bernal}, \citenamefont
  {Rahighi}, \citenamefont {Rusek}, \citenamefont {Smirnov}, \citenamefont
  {Tengblad},\ and\ \citenamefont {Ziman}}]{j.nuclphysa.2007.05.012}%
  \BibitemOpen
  \bibfield  {author} {\bibinfo {author} {\bibfnamefont {D.}~\bibnamefont
  {Escrig}}, \bibinfo {author} {\bibfnamefont {A.~M.}\ \bibnamefont
  {S\'{a}nchez-Ben\'{i}tez}}, \bibinfo {author} {\bibfnamefont {A.~M.}\
  \bibnamefont {Moro}}, \bibinfo {author} {\bibfnamefont {M.~A.~G.}\
  \bibnamefont {Alvarez}}, \bibinfo {author} {\bibfnamefont {M.~V.}\
  \bibnamefont {Andr\'{e}s}}, \bibinfo {author} {\bibfnamefont
  {C.}~\bibnamefont {Angulo}}, \bibinfo {author} {\bibfnamefont {M.~J.~G.}\
  \bibnamefont {Borge}}, \bibinfo {author} {\bibfnamefont {J.}~\bibnamefont
  {Cabrera}}, \bibinfo {author} {\bibfnamefont {S.}~\bibnamefont {Cherubini}},
  \bibinfo {author} {\bibfnamefont {P.}~\bibnamefont {Demaret}}, \bibinfo
  {author} {\bibfnamefont {J.~M.}\ \bibnamefont {Espino}}, \bibinfo {author}
  {\bibfnamefont {P.}~\bibnamefont {Figuera}}, \bibinfo {author} {\bibfnamefont
  {M.}~\bibnamefont {Freer}}, \bibinfo {author} {\bibfnamefont {J.~E.}\
  \bibnamefont {Garc\'{i}a-Ramos}}, \bibinfo {author} {\bibfnamefont
  {J.}~\bibnamefont {G\'{o}mez-Camacho}}, \bibinfo {author} {\bibfnamefont
  {M.}~\bibnamefont {Gulino}}, \bibinfo {author} {\bibfnamefont {O.~R.}\
  \bibnamefont {Kakuee}}, \bibinfo {author} {\bibfnamefont {I.}~\bibnamefont
  {Martel}}, \bibinfo {author} {\bibfnamefont {C.}~\bibnamefont {Metelko}},
  \bibinfo {author} {\bibfnamefont {F.}~\bibnamefont {P\'{e}rez-Bernal}},
  \bibinfo {author} {\bibfnamefont {J.}~\bibnamefont {Rahighi}}, \bibinfo
  {author} {\bibfnamefont {K.}~\bibnamefont {Rusek}}, \bibinfo {author}
  {\bibfnamefont {D.}~\bibnamefont {Smirnov}}, \bibinfo {author} {\bibfnamefont
  {O.}~\bibnamefont {Tengblad}}, \ and\ \bibinfo {author} {\bibfnamefont
  {V.}~\bibnamefont {Ziman}},\ }\href {\doibase
  10.1016/j.nuclphysa.2007.05.012} {\bibfield  {journal} {\bibinfo  {journal}
  {Nucl. Phys. A}\ }\textbf {\bibinfo {volume} {792}},\ \bibinfo {pages} {2}
  (\bibinfo {year} {2007})}\BibitemShut {NoStop}%
\bibitem [{\citenamefont {S\'{a}nchez-Ben\'{i}tez}\ \emph
  {et~al.}(2008)\citenamefont {S\'{a}nchez-Ben\'{i}tez}, \citenamefont
  {Escrig}, \citenamefont {Alvarez}, \citenamefont {Andr\'{e}s}, \citenamefont
  {Angulo}, \citenamefont {Borge}, \citenamefont {Cabrera}, \citenamefont
  {Cherubini}, \citenamefont {Demaret}, \citenamefont {Espino}, \citenamefont
  {Figuera}, \citenamefont {Freer}, \citenamefont {Garc\'{i}a-Ramos},
  \citenamefont {G\'{o}mez-Camacho}, \citenamefont {Gulino}, \citenamefont
  {Kakuee}, \citenamefont {Martel}, \citenamefont {Metelko}, \citenamefont
  {Moro}, \citenamefont {P\'{e}rez-Bernal}, \citenamefont {Rahighi},
  \citenamefont {Rusek}, \citenamefont {Smirnov}, \citenamefont {Tengblad},\
  and\ \citenamefont {Ziman}}]{j.nuclphysa.2008.01.030}%
  \BibitemOpen
  \bibfield  {author} {\bibinfo {author} {\bibfnamefont {A.~M.}\ \bibnamefont
  {S\'{a}nchez-Ben\'{i}tez}}, \bibinfo {author} {\bibfnamefont
  {D.}~\bibnamefont {Escrig}}, \bibinfo {author} {\bibfnamefont {M.~A.~G.}\
  \bibnamefont {Alvarez}}, \bibinfo {author} {\bibfnamefont {M.~V.}\
  \bibnamefont {Andr\'{e}s}}, \bibinfo {author} {\bibfnamefont
  {C.}~\bibnamefont {Angulo}}, \bibinfo {author} {\bibfnamefont {M.~J.~G.}\
  \bibnamefont {Borge}}, \bibinfo {author} {\bibfnamefont {J.}~\bibnamefont
  {Cabrera}}, \bibinfo {author} {\bibfnamefont {S.}~\bibnamefont {Cherubini}},
  \bibinfo {author} {\bibfnamefont {P.}~\bibnamefont {Demaret}}, \bibinfo
  {author} {\bibfnamefont {J.~M.}\ \bibnamefont {Espino}}, \bibinfo {author}
  {\bibfnamefont {P.}~\bibnamefont {Figuera}}, \bibinfo {author} {\bibfnamefont
  {M.}~\bibnamefont {Freer}}, \bibinfo {author} {\bibfnamefont {J.~E.}\
  \bibnamefont {Garc\'{i}a-Ramos}}, \bibinfo {author} {\bibfnamefont
  {J.}~\bibnamefont {G\'{o}mez-Camacho}}, \bibinfo {author} {\bibfnamefont
  {M.}~\bibnamefont {Gulino}}, \bibinfo {author} {\bibfnamefont {O.~R.}\
  \bibnamefont {Kakuee}}, \bibinfo {author} {\bibfnamefont {I.}~\bibnamefont
  {Martel}}, \bibinfo {author} {\bibfnamefont {C.}~\bibnamefont {Metelko}},
  \bibinfo {author} {\bibfnamefont {A.~M.}\ \bibnamefont {Moro}}, \bibinfo
  {author} {\bibfnamefont {F.}~\bibnamefont {P\'{e}rez-Bernal}}, \bibinfo
  {author} {\bibfnamefont {J.}~\bibnamefont {Rahighi}}, \bibinfo {author}
  {\bibfnamefont {K.}~\bibnamefont {Rusek}}, \bibinfo {author} {\bibfnamefont
  {D.}~\bibnamefont {Smirnov}}, \bibinfo {author} {\bibfnamefont
  {O.}~\bibnamefont {Tengblad}}, \ and\ \bibinfo {author} {\bibfnamefont
  {V.}~\bibnamefont {Ziman}},\ }\href {\doibase
  10.1016/j.nuclphysa.2008.01.030} {\bibfield  {journal} {\bibinfo  {journal}
  {Nucl. Phys. A}\ }\textbf {\bibinfo {volume} {803}},\ \bibinfo {pages} {30}
  (\bibinfo {year} {2008})}\BibitemShut {NoStop}%
\bibitem [{\citenamefont {Acosta}\ \emph {et~al.}(2009)\citenamefont {Acosta},
  \citenamefont {Alvarez}, \citenamefont {Andr{\'e}s}, \citenamefont {Borge},
  \citenamefont {Cort{\'e}s}, \citenamefont {Espino}, \citenamefont {Galaviz},
  \citenamefont {G{\'o}mez-Camacho}, \citenamefont {Maira}, \citenamefont
  {Martel}, \citenamefont {Moro}, \citenamefont {Mukha}, \citenamefont
  {P{\'e}rez-Bernal}, \citenamefont {Reillo}, \citenamefont {Rodr{\'i}guez},
  \citenamefont {Rusek}, \citenamefont {S{\'a}nchez-Ben{\'i}tez},\ and\
  \citenamefont {Tengblad}}]{i2009-10822-6}%
  \BibitemOpen
  \bibfield  {author} {\bibinfo {author} {\bibfnamefont {L.}~\bibnamefont
  {Acosta}}, \bibinfo {author} {\bibfnamefont {M.~A.~G.}\ \bibnamefont
  {Alvarez}}, \bibinfo {author} {\bibfnamefont {M.~V.}\ \bibnamefont
  {Andr{\'e}s}}, \bibinfo {author} {\bibfnamefont {M.~J.~G.}\ \bibnamefont
  {Borge}}, \bibinfo {author} {\bibfnamefont {M.}~\bibnamefont {Cort{\'e}s}},
  \bibinfo {author} {\bibfnamefont {J.~M.}\ \bibnamefont {Espino}}, \bibinfo
  {author} {\bibfnamefont {D.}~\bibnamefont {Galaviz}}, \bibinfo {author}
  {\bibfnamefont {J.}~\bibnamefont {G{\'o}mez-Camacho}}, \bibinfo {author}
  {\bibfnamefont {A.}~\bibnamefont {Maira}}, \bibinfo {author} {\bibfnamefont
  {I.}~\bibnamefont {Martel}}, \bibinfo {author} {\bibfnamefont {A.~M.}\
  \bibnamefont {Moro}}, \bibinfo {author} {\bibfnamefont {I.}~\bibnamefont
  {Mukha}}, \bibinfo {author} {\bibfnamefont {F.}~\bibnamefont
  {P{\'e}rez-Bernal}}, \bibinfo {author} {\bibfnamefont {E.}~\bibnamefont
  {Reillo}}, \bibinfo {author} {\bibfnamefont {D.}~\bibnamefont
  {Rodr{\'i}guez}}, \bibinfo {author} {\bibfnamefont {K.}~\bibnamefont
  {Rusek}}, \bibinfo {author} {\bibfnamefont {A.~M.}\ \bibnamefont
  {S{\'a}nchez-Ben{\'i}tez}}, \ and\ \bibinfo {author} {\bibfnamefont
  {O.}~\bibnamefont {Tengblad}},\ }\href {\doibase 10.1140/epja/i2009-10822-6}
  {\bibfield  {journal} {\bibinfo  {journal} {Eur. Phys. J. A}\ }\textbf
  {\bibinfo {volume} {42}},\ \bibinfo {pages} {461} (\bibinfo {year}
  {2009})}\BibitemShut {NoStop}%
\bibitem [{\citenamefont {Acosta}\ \emph {et~al.}(2011)\citenamefont {Acosta},
  \citenamefont {S\'anchez-Ben\'{\i}tez}, \citenamefont {G\'omez},
  \citenamefont {Martel}, \citenamefont {P\'erez-Bernal}, \citenamefont
  {Pizarro}, \citenamefont {Rodr\'{\i}guez-Quintero}, \citenamefont {Rusek},
  \citenamefont {Alvarez}, \citenamefont {Andr\'es}, \citenamefont {Espino},
  \citenamefont {Fern\'andez-Garc\'{\i}a}, \citenamefont {G\'omez-Camacho},
  \citenamefont {Moro}, \citenamefont {Angulo}, \citenamefont {Cabrera},
  \citenamefont {Casarejos}, \citenamefont {Demaret}, \citenamefont {Borge},
  \citenamefont {Escrig}, \citenamefont {Tengblad}, \citenamefont {Cherubini},
  \citenamefont {Figuera}, \citenamefont {Gulino}, \citenamefont {Freer},
  \citenamefont {Metelko}, \citenamefont {Ziman}, \citenamefont {Raabe},
  \citenamefont {Mukha}, \citenamefont {Smirnov}, \citenamefont {Kakuee},\ and\
  \citenamefont {Rahighi}}]{PhysRevC.84.044604}%
  \BibitemOpen
  \bibfield  {author} {\bibinfo {author} {\bibfnamefont {L.}~\bibnamefont
  {Acosta}}, \bibinfo {author} {\bibfnamefont {A.~M.}\ \bibnamefont
  {S\'anchez-Ben\'{\i}tez}}, \bibinfo {author} {\bibfnamefont {M.~E.}\
  \bibnamefont {G\'omez}}, \bibinfo {author} {\bibfnamefont {I.}~\bibnamefont
  {Martel}}, \bibinfo {author} {\bibfnamefont {F.}~\bibnamefont
  {P\'erez-Bernal}}, \bibinfo {author} {\bibfnamefont {F.}~\bibnamefont
  {Pizarro}}, \bibinfo {author} {\bibfnamefont {J.}~\bibnamefont
  {Rodr\'{\i}guez-Quintero}}, \bibinfo {author} {\bibfnamefont
  {K.}~\bibnamefont {Rusek}}, \bibinfo {author} {\bibfnamefont {M.~A.~G.}\
  \bibnamefont {Alvarez}}, \bibinfo {author} {\bibfnamefont {M.~V.}\
  \bibnamefont {Andr\'es}}, \bibinfo {author} {\bibfnamefont {J.~M.}\
  \bibnamefont {Espino}}, \bibinfo {author} {\bibfnamefont {J.~P.}\
  \bibnamefont {Fern\'andez-Garc\'{\i}a}}, \bibinfo {author} {\bibfnamefont
  {J.}~\bibnamefont {G\'omez-Camacho}}, \bibinfo {author} {\bibfnamefont
  {A.~M.}\ \bibnamefont {Moro}}, \bibinfo {author} {\bibfnamefont
  {C.}~\bibnamefont {Angulo}}, \bibinfo {author} {\bibfnamefont
  {J.}~\bibnamefont {Cabrera}}, \bibinfo {author} {\bibfnamefont
  {E.}~\bibnamefont {Casarejos}}, \bibinfo {author} {\bibfnamefont
  {P.}~\bibnamefont {Demaret}}, \bibinfo {author} {\bibfnamefont {M.~J.~G.}\
  \bibnamefont {Borge}}, \bibinfo {author} {\bibfnamefont {D.}~\bibnamefont
  {Escrig}}, \bibinfo {author} {\bibfnamefont {O.}~\bibnamefont {Tengblad}},
  \bibinfo {author} {\bibfnamefont {S.}~\bibnamefont {Cherubini}}, \bibinfo
  {author} {\bibfnamefont {P.}~\bibnamefont {Figuera}}, \bibinfo {author}
  {\bibfnamefont {M.}~\bibnamefont {Gulino}}, \bibinfo {author} {\bibfnamefont
  {M.}~\bibnamefont {Freer}}, \bibinfo {author} {\bibfnamefont
  {C.}~\bibnamefont {Metelko}}, \bibinfo {author} {\bibfnamefont
  {V.}~\bibnamefont {Ziman}}, \bibinfo {author} {\bibfnamefont
  {R.}~\bibnamefont {Raabe}}, \bibinfo {author} {\bibfnamefont
  {I.}~\bibnamefont {Mukha}}, \bibinfo {author} {\bibfnamefont
  {D.}~\bibnamefont {Smirnov}}, \bibinfo {author} {\bibfnamefont {O.~R.}\
  \bibnamefont {Kakuee}}, \ and\ \bibinfo {author} {\bibfnamefont
  {J.}~\bibnamefont {Rahighi}},\ }\href {\doibase 10.1103/PhysRevC.84.044604}
  {\bibfield  {journal} {\bibinfo  {journal} {Phys. Rev. C}\ }\textbf {\bibinfo
  {volume} {84}},\ \bibinfo {pages} {044604} (\bibinfo {year}
  {2011})}\BibitemShut {NoStop}%
\bibitem [{\citenamefont {Rafiei}\ \emph {et~al.}(2010)\citenamefont {Rafiei},
  \citenamefont {du~Rietz}, \citenamefont {Luong}, \citenamefont {Hinde},
  \citenamefont {Dasgupta}, \citenamefont {Evers},\ and\ \citenamefont
  {Diaz-Torres}}]{PhysRevC.81.024601}%
  \BibitemOpen
  \bibfield  {author} {\bibinfo {author} {\bibfnamefont {R.}~\bibnamefont
  {Rafiei}}, \bibinfo {author} {\bibfnamefont {R.}~\bibnamefont {du~Rietz}},
  \bibinfo {author} {\bibfnamefont {D.~H.}\ \bibnamefont {Luong}}, \bibinfo
  {author} {\bibfnamefont {D.~J.}\ \bibnamefont {Hinde}}, \bibinfo {author}
  {\bibfnamefont {M.}~\bibnamefont {Dasgupta}}, \bibinfo {author}
  {\bibfnamefont {M.}~\bibnamefont {Evers}}, \ and\ \bibinfo {author}
  {\bibfnamefont {A.}~\bibnamefont {Diaz-Torres}},\ }\href {\doibase
  10.1103/PhysRevC.81.024601} {\bibfield  {journal} {\bibinfo  {journal} {Phys.
  Rev. C}\ }\textbf {\bibinfo {volume} {81}},\ \bibinfo {pages} {024601}
  (\bibinfo {year} {2010})}\BibitemShut {NoStop}%
\bibitem [{\citenamefont {Luong}\ \emph {et~al.}(2011)\citenamefont {Luong},
  \citenamefont {Dasgupta}, \citenamefont {Hinde}, \citenamefont {du~Rietz},
  \citenamefont {Rafiei}, \citenamefont {Lin}, \citenamefont {Evers},\ and\
  \citenamefont {D\'{i}az-Torres}}]{j.physletb.2010.11.007}%
  \BibitemOpen
  \bibfield  {author} {\bibinfo {author} {\bibfnamefont {D.}~\bibnamefont
  {Luong}}, \bibinfo {author} {\bibfnamefont {M.}~\bibnamefont {Dasgupta}},
  \bibinfo {author} {\bibfnamefont {D.~J.}\ \bibnamefont {Hinde}}, \bibinfo
  {author} {\bibfnamefont {R.}~\bibnamefont {du~Rietz}}, \bibinfo {author}
  {\bibfnamefont {R.}~\bibnamefont {Rafiei}}, \bibinfo {author} {\bibfnamefont
  {C.~J.}\ \bibnamefont {Lin}}, \bibinfo {author} {\bibfnamefont
  {M.}~\bibnamefont {Evers}}, \ and\ \bibinfo {author} {\bibfnamefont
  {A.}~\bibnamefont {D\'{i}az-Torres}},\ }\href {\doibase
  10.1016/j.physletb.2010.11.007} {\bibfield  {journal} {\bibinfo  {journal}
  {Phys. Lett. B}\ }\textbf {\bibinfo {volume} {695}},\ \bibinfo {pages} {105}
  (\bibinfo {year} {2011})}\BibitemShut {NoStop}%
\bibitem [{\citenamefont {Fern\'andez-Garc\'{\i}a}\ \emph
  {et~al.}(2013)\citenamefont {Fern\'andez-Garc\'{\i}a}, \citenamefont
  {Cubero}, \citenamefont {Rodr\'{\i}guez-Gallardo}, \citenamefont {Acosta},
  \citenamefont {Alcorta}, \citenamefont {Alvarez}, \citenamefont {Borge},
  \citenamefont {Buchmann}, \citenamefont {Diget}, \citenamefont {Falou},
  \citenamefont {Fulton}, \citenamefont {Fynbo}, \citenamefont {Galaviz},
  \citenamefont {G\'omez-Camacho}, \citenamefont {Kanungo}, \citenamefont
  {Lay}, \citenamefont {Madurga}, \citenamefont {Martel}, \citenamefont {Moro},
  \citenamefont {Mukha}, \citenamefont {Nilsson}, \citenamefont
  {S\'anchez-Ben\'{\i}tez}, \citenamefont {Shotter}, \citenamefont {Tengblad},\
  and\ \citenamefont {Walden}}]{PhysRevLett.110.142701}%
  \BibitemOpen
  \bibfield  {author} {\bibinfo {author} {\bibfnamefont {J.~P.}\ \bibnamefont
  {Fern\'andez-Garc\'{\i}a}}, \bibinfo {author} {\bibfnamefont
  {M.}~\bibnamefont {Cubero}}, \bibinfo {author} {\bibfnamefont
  {M.}~\bibnamefont {Rodr\'{\i}guez-Gallardo}}, \bibinfo {author}
  {\bibfnamefont {L.}~\bibnamefont {Acosta}}, \bibinfo {author} {\bibfnamefont
  {M.}~\bibnamefont {Alcorta}}, \bibinfo {author} {\bibfnamefont {M.~A.~G.}\
  \bibnamefont {Alvarez}}, \bibinfo {author} {\bibfnamefont {M.~J.~G.}\
  \bibnamefont {Borge}}, \bibinfo {author} {\bibfnamefont {L.}~\bibnamefont
  {Buchmann}}, \bibinfo {author} {\bibfnamefont {C.~A.}\ \bibnamefont {Diget}},
  \bibinfo {author} {\bibfnamefont {H.~A.}\ \bibnamefont {Falou}}, \bibinfo
  {author} {\bibfnamefont {B.~R.}\ \bibnamefont {Fulton}}, \bibinfo {author}
  {\bibfnamefont {H.~O.~U.}\ \bibnamefont {Fynbo}}, \bibinfo {author}
  {\bibfnamefont {D.}~\bibnamefont {Galaviz}}, \bibinfo {author} {\bibfnamefont
  {J.}~\bibnamefont {G\'omez-Camacho}}, \bibinfo {author} {\bibfnamefont
  {R.}~\bibnamefont {Kanungo}}, \bibinfo {author} {\bibfnamefont {J.~A.}\
  \bibnamefont {Lay}}, \bibinfo {author} {\bibfnamefont {M.}~\bibnamefont
  {Madurga}}, \bibinfo {author} {\bibfnamefont {I.}~\bibnamefont {Martel}},
  \bibinfo {author} {\bibfnamefont {A.~M.}\ \bibnamefont {Moro}}, \bibinfo
  {author} {\bibfnamefont {I.}~\bibnamefont {Mukha}}, \bibinfo {author}
  {\bibfnamefont {T.}~\bibnamefont {Nilsson}}, \bibinfo {author} {\bibfnamefont
  {A.~M.}\ \bibnamefont {S\'anchez-Ben\'{\i}tez}}, \bibinfo {author}
  {\bibfnamefont {A.}~\bibnamefont {Shotter}}, \bibinfo {author} {\bibfnamefont
  {O.}~\bibnamefont {Tengblad}}, \ and\ \bibinfo {author} {\bibfnamefont
  {P.}~\bibnamefont {Walden}},\ }\href {\doibase
  10.1103/PhysRevLett.110.142701} {\bibfield  {journal} {\bibinfo  {journal}
  {Phys. Rev. Lett.}\ }\textbf {\bibinfo {volume} {110}},\ \bibinfo {pages}
  {142701} (\bibinfo {year} {2013})}\BibitemShut {NoStop}%
\bibitem [{\citenamefont {Kalkal}\ \emph {et~al.}(2016)\citenamefont {Kalkal},
  \citenamefont {Simpson}, \citenamefont {Luong}, \citenamefont {Cook},
  \citenamefont {Dasgupta}, \citenamefont {Hinde}, \citenamefont {Carter},
  \citenamefont {Jeung}, \citenamefont {Mohanto}, \citenamefont {Palshetkar},
  \citenamefont {Prasad}, \citenamefont {Rafferty}, \citenamefont {Simenel},
  \citenamefont {Vo-Phuoc}, \citenamefont {Williams}, \citenamefont {Gasques},
  \citenamefont {Gomes},\ and\ \citenamefont {Linares}}]{PhysRevC.93.044605}%
  \BibitemOpen
  \bibfield  {author} {\bibinfo {author} {\bibfnamefont {S.}~\bibnamefont
  {Kalkal}}, \bibinfo {author} {\bibfnamefont {E.~C.}\ \bibnamefont {Simpson}},
  \bibinfo {author} {\bibfnamefont {D.~H.}\ \bibnamefont {Luong}}, \bibinfo
  {author} {\bibfnamefont {K.~J.}\ \bibnamefont {Cook}}, \bibinfo {author}
  {\bibfnamefont {M.}~\bibnamefont {Dasgupta}}, \bibinfo {author}
  {\bibfnamefont {D.~J.}\ \bibnamefont {Hinde}}, \bibinfo {author}
  {\bibfnamefont {I.~P.}\ \bibnamefont {Carter}}, \bibinfo {author}
  {\bibfnamefont {D.~Y.}\ \bibnamefont {Jeung}}, \bibinfo {author}
  {\bibfnamefont {G.}~\bibnamefont {Mohanto}}, \bibinfo {author} {\bibfnamefont
  {C.~S.}\ \bibnamefont {Palshetkar}}, \bibinfo {author} {\bibfnamefont
  {E.}~\bibnamefont {Prasad}}, \bibinfo {author} {\bibfnamefont {D.~C.}\
  \bibnamefont {Rafferty}}, \bibinfo {author} {\bibfnamefont {C.}~\bibnamefont
  {Simenel}}, \bibinfo {author} {\bibfnamefont {K.}~\bibnamefont {Vo-Phuoc}},
  \bibinfo {author} {\bibfnamefont {E.}~\bibnamefont {Williams}}, \bibinfo
  {author} {\bibfnamefont {L.~R.}\ \bibnamefont {Gasques}}, \bibinfo {author}
  {\bibfnamefont {P.~R.~S.}\ \bibnamefont {Gomes}}, \ and\ \bibinfo {author}
  {\bibfnamefont {R.}~\bibnamefont {Linares}},\ }\href {\doibase
  10.1103/PhysRevC.93.044605} {\bibfield  {journal} {\bibinfo  {journal} {Phys.
  Rev. C}\ }\textbf {\bibinfo {volume} {93}},\ \bibinfo {pages} {044605}
  (\bibinfo {year} {2016})}\BibitemShut {NoStop}%
\bibitem [{\citenamefont {Pesudo}\ \emph {et~al.}(2017)\citenamefont {Pesudo},
  \citenamefont {Borge}, \citenamefont {Moro}, \citenamefont {Lay},
  \citenamefont {N\'acher}, \citenamefont {G\'omez-Camacho}, \citenamefont
  {Tengblad}, \citenamefont {Acosta}, \citenamefont {Alcorta}, \citenamefont
  {Alvarez}, \citenamefont {Andreoiu}, \citenamefont {Bender}, \citenamefont
  {Braid}, \citenamefont {Cubero}, \citenamefont {Di~Pietro}, \citenamefont
  {Fern\'andez-Garc\'{\i}a}, \citenamefont {Figuera}, \citenamefont
  {Fisichella}, \citenamefont {Fulton}, \citenamefont {Garnsworthy},
  \citenamefont {Hackman}, \citenamefont {Hager}, \citenamefont {Kirsebom},
  \citenamefont {Kuhn}, \citenamefont {Lattuada}, \citenamefont
  {Marqu\'{\i}nez-Dur\'an}, \citenamefont {Martel}, \citenamefont {Miller},
  \citenamefont {Moukaddam}, \citenamefont {O'Malley}, \citenamefont {Perea},
  \citenamefont {Rajabali}, \citenamefont {S\'anchez-Ben\'{\i}tez},
  \citenamefont {Sarazin}, \citenamefont {Scuderi}, \citenamefont {Svensson},
  \citenamefont {Unsworth},\ and\ \citenamefont
  {Wang}}]{PhysRevLett.118.152502}%
  \BibitemOpen
  \bibfield  {author} {\bibinfo {author} {\bibfnamefont {V.}~\bibnamefont
  {Pesudo}}, \bibinfo {author} {\bibfnamefont {M.~J.~G.}\ \bibnamefont
  {Borge}}, \bibinfo {author} {\bibfnamefont {A.~M.}\ \bibnamefont {Moro}},
  \bibinfo {author} {\bibfnamefont {J.~A.}\ \bibnamefont {Lay}}, \bibinfo
  {author} {\bibfnamefont {E.}~\bibnamefont {N\'acher}}, \bibinfo {author}
  {\bibfnamefont {J.}~\bibnamefont {G\'omez-Camacho}}, \bibinfo {author}
  {\bibfnamefont {O.}~\bibnamefont {Tengblad}}, \bibinfo {author}
  {\bibfnamefont {L.}~\bibnamefont {Acosta}}, \bibinfo {author} {\bibfnamefont
  {M.}~\bibnamefont {Alcorta}}, \bibinfo {author} {\bibfnamefont {M.~A.~G.}\
  \bibnamefont {Alvarez}}, \bibinfo {author} {\bibfnamefont {C.}~\bibnamefont
  {Andreoiu}}, \bibinfo {author} {\bibfnamefont {P.~C.}\ \bibnamefont
  {Bender}}, \bibinfo {author} {\bibfnamefont {R.}~\bibnamefont {Braid}},
  \bibinfo {author} {\bibfnamefont {M.}~\bibnamefont {Cubero}}, \bibinfo
  {author} {\bibfnamefont {A.}~\bibnamefont {Di~Pietro}}, \bibinfo {author}
  {\bibfnamefont {J.~P.}\ \bibnamefont {Fern\'andez-Garc\'{\i}a}}, \bibinfo
  {author} {\bibfnamefont {P.}~\bibnamefont {Figuera}}, \bibinfo {author}
  {\bibfnamefont {M.}~\bibnamefont {Fisichella}}, \bibinfo {author}
  {\bibfnamefont {B.~R.}\ \bibnamefont {Fulton}}, \bibinfo {author}
  {\bibfnamefont {A.~B.}\ \bibnamefont {Garnsworthy}}, \bibinfo {author}
  {\bibfnamefont {G.}~\bibnamefont {Hackman}}, \bibinfo {author} {\bibfnamefont
  {U.}~\bibnamefont {Hager}}, \bibinfo {author} {\bibfnamefont {O.~S.}\
  \bibnamefont {Kirsebom}}, \bibinfo {author} {\bibfnamefont {K.}~\bibnamefont
  {Kuhn}}, \bibinfo {author} {\bibfnamefont {M.}~\bibnamefont {Lattuada}},
  \bibinfo {author} {\bibfnamefont {G.}~\bibnamefont {Marqu\'{\i}nez-Dur\'an}},
  \bibinfo {author} {\bibfnamefont {I.}~\bibnamefont {Martel}}, \bibinfo
  {author} {\bibfnamefont {D.}~\bibnamefont {Miller}}, \bibinfo {author}
  {\bibfnamefont {M.}~\bibnamefont {Moukaddam}}, \bibinfo {author}
  {\bibfnamefont {P.~D.}\ \bibnamefont {O'Malley}}, \bibinfo {author}
  {\bibfnamefont {A.}~\bibnamefont {Perea}}, \bibinfo {author} {\bibfnamefont
  {M.~M.}\ \bibnamefont {Rajabali}}, \bibinfo {author} {\bibfnamefont {A.~M.}\
  \bibnamefont {S\'anchez-Ben\'{\i}tez}}, \bibinfo {author} {\bibfnamefont
  {F.}~\bibnamefont {Sarazin}}, \bibinfo {author} {\bibfnamefont
  {V.}~\bibnamefont {Scuderi}}, \bibinfo {author} {\bibfnamefont {C.~E.}\
  \bibnamefont {Svensson}}, \bibinfo {author} {\bibfnamefont {C.}~\bibnamefont
  {Unsworth}}, \ and\ \bibinfo {author} {\bibfnamefont {Z.~M.}\ \bibnamefont
  {Wang}},\ }\href {\doibase 10.1103/PhysRevLett.118.152502} {\bibfield
  {journal} {\bibinfo  {journal} {Phys. Rev. Lett.}\ }\textbf {\bibinfo
  {volume} {118}},\ \bibinfo {pages} {152502} (\bibinfo {year}
  {2017})}\BibitemShut {NoStop}%
\bibitem [{\citenamefont {Fern\'andez-Garc\'{\i}a}\ \emph
  {et~al.}(2015)\citenamefont {Fern\'andez-Garc\'{\i}a}, \citenamefont
  {Alvarez},\ and\ \citenamefont {Chamon}}]{PhysRevC.92.014604}%
  \BibitemOpen
  \bibfield  {author} {\bibinfo {author} {\bibfnamefont {J.~P.}\ \bibnamefont
  {Fern\'andez-Garc\'{\i}a}}, \bibinfo {author} {\bibfnamefont {M.~A.~G.}\
  \bibnamefont {Alvarez}}, \ and\ \bibinfo {author} {\bibfnamefont {L.~C.}\
  \bibnamefont {Chamon}},\ }\href {\doibase 10.1103/PhysRevC.92.014604}
  {\bibfield  {journal} {\bibinfo  {journal} {Phys. Rev. C}\ }\textbf {\bibinfo
  {volume} {92}},\ \bibinfo {pages} {014604} (\bibinfo {year}
  {2015})}\BibitemShut {NoStop}%
\bibitem [{\citenamefont {Kelley}\ \emph {et~al.}(2017)\citenamefont {Kelley},
  \citenamefont {Purcell},\ and\ \citenamefont {Sheu}}]{KELLEY201771}%
  \BibitemOpen
  \bibfield  {author} {\bibinfo {author} {\bibfnamefont {J.}~\bibnamefont
  {Kelley}}, \bibinfo {author} {\bibfnamefont {J.}~\bibnamefont {Purcell}}, \
  and\ \bibinfo {author} {\bibfnamefont {C.}~\bibnamefont {Sheu}},\ }\href
  {\doibase https://doi.org/10.1016/j.nuclphysa.2017.07.015} {\bibfield
  {journal} {\bibinfo  {journal} {Nuclear Physics A}\ }\textbf {\bibinfo
  {volume} {968}},\ \bibinfo {pages} {71} (\bibinfo {year} {2017})}\BibitemShut
  {NoStop}%
\bibitem [{\citenamefont {Tilley}\ \emph {et~al.}(2004)\citenamefont {Tilley},
  \citenamefont {Kelley}, \citenamefont {Godwin}, \citenamefont {Millener},
  \citenamefont {Purcell}, \citenamefont {Sheu},\ and\ \citenamefont
  {Weller}}]{j.nuclphysa.2004.09.059}%
  \BibitemOpen
  \bibfield  {author} {\bibinfo {author} {\bibfnamefont {D.~R.}\ \bibnamefont
  {Tilley}}, \bibinfo {author} {\bibfnamefont {J.~H.}\ \bibnamefont {Kelley}},
  \bibinfo {author} {\bibfnamefont {J.~L.}\ \bibnamefont {Godwin}}, \bibinfo
  {author} {\bibfnamefont {D.~J.}\ \bibnamefont {Millener}}, \bibinfo {author}
  {\bibfnamefont {J.~E.}\ \bibnamefont {Purcell}}, \bibinfo {author}
  {\bibfnamefont {C.~G.}\ \bibnamefont {Sheu}}, \ and\ \bibinfo {author}
  {\bibfnamefont {H.~R.}\ \bibnamefont {Weller}},\ }\href {\doibase
  10.1016/j.nuclphysa.2004.09.059} {\bibfield  {journal} {\bibinfo  {journal}
  {Nucl. Phys. A}\ }\textbf {\bibinfo {volume} {745}},\ \bibinfo {pages} {155}
  (\bibinfo {year} {2004})}\BibitemShut {NoStop}%
\bibitem [{\citenamefont {Tilley}\ \emph {et~al.}(2002)\citenamefont {Tilley},
  \citenamefont {Cheves}, \citenamefont {Godwin}, \citenamefont {Hale},
  \citenamefont {Hofmann}, \citenamefont {Kelley}, \citenamefont {Sheu},\ and\
  \citenamefont {Weller}}]{S0375-9474(02)00597-3}%
  \BibitemOpen
  \bibfield  {author} {\bibinfo {author} {\bibfnamefont {D.~R.}\ \bibnamefont
  {Tilley}}, \bibinfo {author} {\bibfnamefont {C.~M.}\ \bibnamefont {Cheves}},
  \bibinfo {author} {\bibfnamefont {J.~L.}\ \bibnamefont {Godwin}}, \bibinfo
  {author} {\bibfnamefont {J.~M.}\ \bibnamefont {Hale}}, \bibinfo {author}
  {\bibfnamefont {H.~M.}\ \bibnamefont {Hofmann}}, \bibinfo {author}
  {\bibfnamefont {J.~H.}\ \bibnamefont {Kelley}}, \bibinfo {author}
  {\bibfnamefont {C.~G.}\ \bibnamefont {Sheu}}, \ and\ \bibinfo {author}
  {\bibfnamefont {H.~R.}\ \bibnamefont {Weller}},\ }\href {\doibase
  10.1016/S0375-9474(02)00597-3} {\bibfield  {journal} {\bibinfo  {journal}
  {Nucl. Phys. A}\ }\textbf {\bibinfo {volume} {708}},\ \bibinfo {pages} {3}
  (\bibinfo {year} {2002})}\BibitemShut {NoStop}%
\bibitem [{\citenamefont {Aguilera}\ \emph {et~al.}(2000)\citenamefont
  {Aguilera}, \citenamefont {Kolata}, \citenamefont {Nunes}, \citenamefont
  {Becchetti}, \citenamefont {DeYoung}, \citenamefont {Goupell}, \citenamefont
  {Guimar\~aes}, \citenamefont {Hughey}, \citenamefont {Lee}, \citenamefont
  {Lizcano}, \citenamefont {Martinez-Quiroz}, \citenamefont {Nowlin},
  \citenamefont {O'Donnell}, \citenamefont {Peaslee}, \citenamefont {Peterson},
  \citenamefont {Santi},\ and\ \citenamefont
  {White-Stevens}}]{PhysRevLett.84.5058}%
  \BibitemOpen
  \bibfield  {author} {\bibinfo {author} {\bibfnamefont {E.~F.}\ \bibnamefont
  {Aguilera}}, \bibinfo {author} {\bibfnamefont {J.~J.}\ \bibnamefont
  {Kolata}}, \bibinfo {author} {\bibfnamefont {F.~M.}\ \bibnamefont {Nunes}},
  \bibinfo {author} {\bibfnamefont {F.~D.}\ \bibnamefont {Becchetti}}, \bibinfo
  {author} {\bibfnamefont {P.~A.}\ \bibnamefont {DeYoung}}, \bibinfo {author}
  {\bibfnamefont {M.}~\bibnamefont {Goupell}}, \bibinfo {author} {\bibfnamefont
  {V.}~\bibnamefont {Guimar\~aes}}, \bibinfo {author} {\bibfnamefont
  {B.}~\bibnamefont {Hughey}}, \bibinfo {author} {\bibfnamefont {M.~Y.}\
  \bibnamefont {Lee}}, \bibinfo {author} {\bibfnamefont {D.}~\bibnamefont
  {Lizcano}}, \bibinfo {author} {\bibfnamefont {E.}~\bibnamefont
  {Martinez-Quiroz}}, \bibinfo {author} {\bibfnamefont {A.}~\bibnamefont
  {Nowlin}}, \bibinfo {author} {\bibfnamefont {T.~W.}\ \bibnamefont
  {O'Donnell}}, \bibinfo {author} {\bibfnamefont {G.~F.}\ \bibnamefont
  {Peaslee}}, \bibinfo {author} {\bibfnamefont {D.}~\bibnamefont {Peterson}},
  \bibinfo {author} {\bibfnamefont {P.}~\bibnamefont {Santi}}, \ and\ \bibinfo
  {author} {\bibfnamefont {R.}~\bibnamefont {White-Stevens}},\ }\href {\doibase
  10.1103/PhysRevLett.84.5058} {\bibfield  {journal} {\bibinfo  {journal}
  {Phys. Rev. Lett.}\ }\textbf {\bibinfo {volume} {84}},\ \bibinfo {pages}
  {5058} (\bibinfo {year} {2000})}\BibitemShut {NoStop}%
\bibitem [{\citenamefont {Di~Pietro}\ \emph {et~al.}(2004)\citenamefont
  {Di~Pietro}, \citenamefont {Figuera}, \citenamefont {Amorini}, \citenamefont
  {Angulo}, \citenamefont {Cardella}, \citenamefont {Cherubini}, \citenamefont
  {Davinson}, \citenamefont {Leanza}, \citenamefont {Lu}, \citenamefont
  {Mahmud}, \citenamefont {Milin}, \citenamefont {Musumarra}, \citenamefont
  {Ninane}, \citenamefont {Papa}, \citenamefont {Pellegriti}, \citenamefont
  {Raabe}, \citenamefont {Rizzo}, \citenamefont {Ruiz}, \citenamefont
  {Shotter}, \citenamefont {Soi\ifmmode~\acute{c}\else \'{c}\fi{}},
  \citenamefont {Tudisco},\ and\ \citenamefont
  {Weissman}}]{PhysRevC.69.044613}%
  \BibitemOpen
  \bibfield  {author} {\bibinfo {author} {\bibfnamefont {A.}~\bibnamefont
  {Di~Pietro}}, \bibinfo {author} {\bibfnamefont {P.}~\bibnamefont {Figuera}},
  \bibinfo {author} {\bibfnamefont {F.}~\bibnamefont {Amorini}}, \bibinfo
  {author} {\bibfnamefont {C.}~\bibnamefont {Angulo}}, \bibinfo {author}
  {\bibfnamefont {G.}~\bibnamefont {Cardella}}, \bibinfo {author}
  {\bibfnamefont {S.}~\bibnamefont {Cherubini}}, \bibinfo {author}
  {\bibfnamefont {T.}~\bibnamefont {Davinson}}, \bibinfo {author}
  {\bibfnamefont {D.}~\bibnamefont {Leanza}}, \bibinfo {author} {\bibfnamefont
  {J.}~\bibnamefont {Lu}}, \bibinfo {author} {\bibfnamefont {H.}~\bibnamefont
  {Mahmud}}, \bibinfo {author} {\bibfnamefont {M.}~\bibnamefont {Milin}},
  \bibinfo {author} {\bibfnamefont {A.}~\bibnamefont {Musumarra}}, \bibinfo
  {author} {\bibfnamefont {A.}~\bibnamefont {Ninane}}, \bibinfo {author}
  {\bibfnamefont {M.}~\bibnamefont {Papa}}, \bibinfo {author} {\bibfnamefont
  {M.~G.}\ \bibnamefont {Pellegriti}}, \bibinfo {author} {\bibfnamefont
  {R.}~\bibnamefont {Raabe}}, \bibinfo {author} {\bibfnamefont
  {F.}~\bibnamefont {Rizzo}}, \bibinfo {author} {\bibfnamefont
  {C.}~\bibnamefont {Ruiz}}, \bibinfo {author} {\bibfnamefont {A.~C.}\
  \bibnamefont {Shotter}}, \bibinfo {author} {\bibfnamefont {N.}~\bibnamefont
  {Soi\ifmmode~\acute{c}\else \'{c}\fi{}}}, \bibinfo {author} {\bibfnamefont
  {S.}~\bibnamefont {Tudisco}}, \ and\ \bibinfo {author} {\bibfnamefont
  {L.}~\bibnamefont {Weissman}},\ }\href {\doibase 10.1103/PhysRevC.69.044613}
  {\bibfield  {journal} {\bibinfo  {journal} {Phys. Rev. C}\ }\textbf {\bibinfo
  {volume} {69}},\ \bibinfo {pages} {044613} (\bibinfo {year}
  {2004})}\BibitemShut {NoStop}%
\bibitem [{\citenamefont {Fern\'andez-Garc\'{\i}a}\ \emph
  {et~al.}(2019)\citenamefont {Fern\'andez-Garc\'{\i}a}, \citenamefont
  {Di~Pietro}, \citenamefont {Figuera}, \citenamefont {G\'omez-Camacho},
  \citenamefont {Lattuada}, \citenamefont {Lei}, \citenamefont {Moro},
  \citenamefont {Rodr\'{\i}guez-Gallardo},\ and\ \citenamefont
  {Scuderi}}]{PhysRevC.99.054605}%
  \BibitemOpen
  \bibfield  {author} {\bibinfo {author} {\bibfnamefont {J.~P.}\ \bibnamefont
  {Fern\'andez-Garc\'{\i}a}}, \bibinfo {author} {\bibfnamefont
  {A.}~\bibnamefont {Di~Pietro}}, \bibinfo {author} {\bibfnamefont
  {P.}~\bibnamefont {Figuera}}, \bibinfo {author} {\bibfnamefont
  {J.}~\bibnamefont {G\'omez-Camacho}}, \bibinfo {author} {\bibfnamefont
  {M.}~\bibnamefont {Lattuada}}, \bibinfo {author} {\bibfnamefont
  {J.}~\bibnamefont {Lei}}, \bibinfo {author} {\bibfnamefont {A.~M.}\
  \bibnamefont {Moro}}, \bibinfo {author} {\bibfnamefont {M.}~\bibnamefont
  {Rodr\'{\i}guez-Gallardo}}, \ and\ \bibinfo {author} {\bibfnamefont
  {V.}~\bibnamefont {Scuderi}},\ }\href {\doibase 10.1103/PhysRevC.99.054605}
  {\bibfield  {journal} {\bibinfo  {journal} {Phys. Rev. C}\ }\textbf {\bibinfo
  {volume} {99}},\ \bibinfo {pages} {054605} (\bibinfo {year}
  {2019})}\BibitemShut {NoStop}%
\bibitem [{\citenamefont {Fern\'andez-Garc\'{\i}a}\ \emph
  {et~al.}(2010{\natexlab{a}})\citenamefont {Fern\'andez-Garc\'{\i}a},
  \citenamefont {Rodr\'iguez-Gallardo}, \citenamefont {Alvarez},\ and\
  \citenamefont {Moro}}]{j.nuclphysa.2010.03.013}%
  \BibitemOpen
  \bibfield  {author} {\bibinfo {author} {\bibfnamefont {J.~P.}\ \bibnamefont
  {Fern\'andez-Garc\'{\i}a}}, \bibinfo {author} {\bibfnamefont
  {M.}~\bibnamefont {Rodr\'iguez-Gallardo}}, \bibinfo {author} {\bibfnamefont
  {M.~A.~G.}\ \bibnamefont {Alvarez}}, \ and\ \bibinfo {author} {\bibfnamefont
  {A.~M.}\ \bibnamefont {Moro}},\ }\href {\doibase
  10.1016/j.nuclphysa.2010.03.013} {\bibfield  {journal} {\bibinfo  {journal}
  {Nucl. Phys. A}\ }\textbf {\bibinfo {volume} {840}},\ \bibinfo {pages} {19}
  (\bibinfo {year} {2010}{\natexlab{a}})}\BibitemShut {NoStop}%
\bibitem [{\citenamefont {Fern\'andez-Garc\'{\i}a}\ \emph
  {et~al.}(2010{\natexlab{b}})\citenamefont {Fern\'andez-Garc\'{\i}a},
  \citenamefont {Alvarez}, \citenamefont {Moro},\ and\ \citenamefont
  {Rodr\'iguez-Gallardo}}]{j.physletb.2010.07.060}%
  \BibitemOpen
  \bibfield  {author} {\bibinfo {author} {\bibfnamefont {J.~P.}\ \bibnamefont
  {Fern\'andez-Garc\'{\i}a}}, \bibinfo {author} {\bibfnamefont {M.~A.~G.}\
  \bibnamefont {Alvarez}}, \bibinfo {author} {\bibfnamefont {A.~M.}\
  \bibnamefont {Moro}}, \ and\ \bibinfo {author} {\bibfnamefont
  {M.}~\bibnamefont {Rodr\'iguez-Gallardo}},\ }\href {\doibase
  10.1016/j.physletb.2010.07.060} {\bibfield  {journal} {\bibinfo  {journal}
  {Phys. Lett. B}\ }\textbf {\bibinfo {volume} {693}},\ \bibinfo {pages} {310}
  (\bibinfo {year} {2010}{\natexlab{b}})}\BibitemShut {NoStop}%
\bibitem [{\citenamefont {Kelley}\ \emph {et~al.}(2012)\citenamefont {Kelley},
  \citenamefont {Kwan}, \citenamefont {Purcell}, \citenamefont {Sheu},\ and\
  \citenamefont {Weller}}]{j.nuclphysa.2012.01.010}%
  \BibitemOpen
  \bibfield  {author} {\bibinfo {author} {\bibfnamefont {J.~H.}\ \bibnamefont
  {Kelley}}, \bibinfo {author} {\bibfnamefont {E.}~\bibnamefont {Kwan}},
  \bibinfo {author} {\bibfnamefont {J.~E.}\ \bibnamefont {Purcell}}, \bibinfo
  {author} {\bibfnamefont {C.~G.}\ \bibnamefont {Sheu}}, \ and\ \bibinfo
  {author} {\bibfnamefont {H.~R.}\ \bibnamefont {Weller}},\ }\href {\doibase
  10.1016/j.nuclphysa.2012.01.010} {\bibfield  {journal} {\bibinfo  {journal}
  {Nucl. Phys. A}\ }\textbf {\bibinfo {volume} {880}},\ \bibinfo {pages} {88}
  (\bibinfo {year} {2012})}\BibitemShut {NoStop}%
\bibitem [{\citenamefont {Kwan}\ \emph {et~al.}(2014)\citenamefont {Kwan},
  \citenamefont {Wu}, \citenamefont {Summers}, \citenamefont {Hackman},
  \citenamefont {Drake}, \citenamefont {Andreoiu}, \citenamefont {Ashley},
  \citenamefont {Ball}, \citenamefont {Bender}, \citenamefont {Boston},
  \citenamefont {Boston}, \citenamefont {Chester}, \citenamefont {Close},
  \citenamefont {Cline}, \citenamefont {Cross}, \citenamefont {Dunlop},
  \citenamefont {Finlay}, \citenamefont {Garnsworthy},\ and\ \citenamefont
  {Wang}}]{j.physletb.2014.03.049}%
  \BibitemOpen
  \bibfield  {author} {\bibinfo {author} {\bibfnamefont {E.}~\bibnamefont
  {Kwan}}, \bibinfo {author} {\bibfnamefont {C.}~\bibnamefont {Wu}}, \bibinfo
  {author} {\bibfnamefont {C.}~\bibnamefont {Summers}}, \bibinfo {author}
  {\bibfnamefont {G.}~\bibnamefont {Hackman}}, \bibinfo {author} {\bibfnamefont
  {T.~E.}\ \bibnamefont {Drake}}, \bibinfo {author} {\bibfnamefont
  {C.}~\bibnamefont {Andreoiu}}, \bibinfo {author} {\bibfnamefont
  {R.}~\bibnamefont {Ashley}}, \bibinfo {author} {\bibfnamefont {G.~C.}\
  \bibnamefont {Ball}}, \bibinfo {author} {\bibfnamefont {P.~C.}\ \bibnamefont
  {Bender}}, \bibinfo {author} {\bibfnamefont {A.~J.}\ \bibnamefont {Boston}},
  \bibinfo {author} {\bibfnamefont {H.~C.}\ \bibnamefont {Boston}}, \bibinfo
  {author} {\bibfnamefont {A.}~\bibnamefont {Chester}}, \bibinfo {author}
  {\bibfnamefont {A.}~\bibnamefont {Close}}, \bibinfo {author} {\bibfnamefont
  {D.}~\bibnamefont {Cline}}, \bibinfo {author} {\bibfnamefont {D.~S.}\
  \bibnamefont {Cross}}, \bibinfo {author} {\bibfnamefont {R.}~\bibnamefont
  {Dunlop}}, \bibinfo {author} {\bibfnamefont {A.}~\bibnamefont {Finlay}},
  \bibinfo {author} {\bibfnamefont {A.~B.}\ \bibnamefont {Garnsworthy}}, \ and\
  \bibinfo {author} {\bibfnamefont {Z.~M.}\ \bibnamefont {Wang}},\ }\href
  {\doibase 10.1016/j.physletb.2014.03.049} {\bibfield  {journal} {\bibinfo
  {journal} {Nucl. Phys. A}\ }\textbf {\bibinfo {volume} {732}},\ \bibinfo
  {pages} {210} (\bibinfo {year} {2014})}\BibitemShut {NoStop}%
\bibitem [{\citenamefont {Smith}\ \emph {et~al.}(2008)\citenamefont {Smith},
  \citenamefont {Brodeur}, \citenamefont {Brunner}, \citenamefont {Ettenauer},
  \citenamefont {Lapierre}, \citenamefont {Ringle}, \citenamefont {Ryjkov},
  \citenamefont {Ames}, \citenamefont {Bricault}, \citenamefont {Drake},
  \citenamefont {Delheij}, \citenamefont {Lunney}, \citenamefont {Sarazin},\
  and\ \citenamefont {Dilling}}]{PhysRevLett.101.202501}%
  \BibitemOpen
  \bibfield  {author} {\bibinfo {author} {\bibfnamefont {M.}~\bibnamefont
  {Smith}}, \bibinfo {author} {\bibfnamefont {M.}~\bibnamefont {Brodeur}},
  \bibinfo {author} {\bibfnamefont {T.}~\bibnamefont {Brunner}}, \bibinfo
  {author} {\bibfnamefont {S.}~\bibnamefont {Ettenauer}}, \bibinfo {author}
  {\bibfnamefont {A.}~\bibnamefont {Lapierre}}, \bibinfo {author}
  {\bibfnamefont {R.}~\bibnamefont {Ringle}}, \bibinfo {author} {\bibfnamefont
  {V.~L.}\ \bibnamefont {Ryjkov}}, \bibinfo {author} {\bibfnamefont
  {F.}~\bibnamefont {Ames}}, \bibinfo {author} {\bibfnamefont {P.}~\bibnamefont
  {Bricault}}, \bibinfo {author} {\bibfnamefont {G.~W.~F.}\ \bibnamefont
  {Drake}}, \bibinfo {author} {\bibfnamefont {P.}~\bibnamefont {Delheij}},
  \bibinfo {author} {\bibfnamefont {D.}~\bibnamefont {Lunney}}, \bibinfo
  {author} {\bibfnamefont {F.}~\bibnamefont {Sarazin}}, \ and\ \bibinfo
  {author} {\bibfnamefont {J.}~\bibnamefont {Dilling}},\ }\href {\doibase
  10.1103/PhysRevLett.101.202501} {\bibfield  {journal} {\bibinfo  {journal}
  {Phys. Rev. Lett.}\ }\textbf {\bibinfo {volume} {101}},\ \bibinfo {pages}
  {202501} (\bibinfo {year} {2008})}\BibitemShut {NoStop}%
\bibitem [{\citenamefont {G\'{o}mez-Camacho}\ \emph {et~al.}(1994)\citenamefont
  {G\'{o}mez-Camacho}, \citenamefont {Andr\'{e}s},\ and\ \citenamefont
  {Nagarajan}}]{0375-9474(94)90820-6}%
  \BibitemOpen
  \bibfield  {author} {\bibinfo {author} {\bibfnamefont {J.}~\bibnamefont
  {G\'{o}mez-Camacho}}, \bibinfo {author} {\bibfnamefont {M.~V.}\ \bibnamefont
  {Andr\'{e}s}}, \ and\ \bibinfo {author} {\bibfnamefont {N.}~\bibnamefont
  {Nagarajan}},\ }\href {\doibase 10.1016/0375-9474(94)90820-6} {\bibfield
  {journal} {\bibinfo  {journal} {Nucl. Phys. A}\ }\textbf {\bibinfo {volume}
  {580}},\ \bibinfo {pages} {156} (\bibinfo {year} {1994})}\BibitemShut
  {NoStop}%
\bibitem [{\citenamefont {Andr\'{e}s}\ \emph {et~al.}(1995)\citenamefont
  {Andr\'{e}s}, \citenamefont {G\'{o}mez-Camacho},\ and\ \citenamefont
  {Nagarajan}}]{0375-9474(94)00765-F}%
  \BibitemOpen
  \bibfield  {author} {\bibinfo {author} {\bibfnamefont {M.~V.}\ \bibnamefont
  {Andr\'{e}s}}, \bibinfo {author} {\bibfnamefont {J.}~\bibnamefont
  {G\'{o}mez-Camacho}}, \ and\ \bibinfo {author} {\bibfnamefont
  {N.}~\bibnamefont {Nagarajan}},\ }\href {\doibase
  10.1016/0375-9474(94)00765-F} {\bibfield  {journal} {\bibinfo  {journal}
  {Nucl. Phys. A}\ }\textbf {\bibinfo {volume} {583}},\ \bibinfo {pages} {817}
  (\bibinfo {year} {1995})}\BibitemShut {NoStop}%
\bibitem [{\citenamefont {Chamon}\ \emph {et~al.}(2002)\citenamefont {Chamon},
  \citenamefont {Carlson}, \citenamefont {Gasques}, \citenamefont {Pereira},
  \citenamefont {De~Conti}, \citenamefont {Alvarez}, \citenamefont {Hussein},
  \citenamefont {C\^andido~Ribeiro}, \citenamefont {Rossi},\ and\ \citenamefont
  {Silva}}]{PhysRevC.66.014610}%
  \BibitemOpen
  \bibfield  {author} {\bibinfo {author} {\bibfnamefont {L.~C.}\ \bibnamefont
  {Chamon}}, \bibinfo {author} {\bibfnamefont {B.~V.}\ \bibnamefont {Carlson}},
  \bibinfo {author} {\bibfnamefont {L.~R.}\ \bibnamefont {Gasques}}, \bibinfo
  {author} {\bibfnamefont {D.}~\bibnamefont {Pereira}}, \bibinfo {author}
  {\bibfnamefont {C.}~\bibnamefont {De~Conti}}, \bibinfo {author}
  {\bibfnamefont {M.~A.~G.}\ \bibnamefont {Alvarez}}, \bibinfo {author}
  {\bibfnamefont {M.~S.}\ \bibnamefont {Hussein}}, \bibinfo {author}
  {\bibfnamefont {M.~A.}\ \bibnamefont {C\^andido~Ribeiro}}, \bibinfo {author}
  {\bibfnamefont {E.~S.}\ \bibnamefont {Rossi}}, \ and\ \bibinfo {author}
  {\bibfnamefont {C.~P.}\ \bibnamefont {Silva}},\ }\href {\doibase
  10.1103/PhysRevC.66.014610} {\bibfield  {journal} {\bibinfo  {journal} {Phys.
  Rev. C}\ }\textbf {\bibinfo {volume} {66}},\ \bibinfo {pages} {014610}
  (\bibinfo {year} {2002})}\BibitemShut {NoStop}%
\bibitem [{\citenamefont {C\^andido~Ribeiro}\ \emph {et~al.}(1997)\citenamefont
  {C\^andido~Ribeiro}, \citenamefont {Chamon}, \citenamefont {Pereira},
  \citenamefont {Hussein},\ and\ \citenamefont
  {Galetti}}]{PhysRevLett.78.3270}%
  \BibitemOpen
  \bibfield  {author} {\bibinfo {author} {\bibfnamefont {M.~A.}\ \bibnamefont
  {C\^andido~Ribeiro}}, \bibinfo {author} {\bibfnamefont {L.~C.}\ \bibnamefont
  {Chamon}}, \bibinfo {author} {\bibfnamefont {D.}~\bibnamefont {Pereira}},
  \bibinfo {author} {\bibfnamefont {M.~S.}\ \bibnamefont {Hussein}}, \ and\
  \bibinfo {author} {\bibfnamefont {D.}~\bibnamefont {Galetti}},\ }\href
  {\doibase 10.1103/PhysRevLett.78.3270} {\bibfield  {journal} {\bibinfo
  {journal} {Phys. Rev. Lett.}\ }\textbf {\bibinfo {volume} {78}},\ \bibinfo
  {pages} {3270} (\bibinfo {year} {1997})}\BibitemShut {NoStop}%
\bibitem [{\citenamefont {Chamon}\ \emph {et~al.}(1997)\citenamefont {Chamon},
  \citenamefont {Pereira}, \citenamefont {Hussein}, \citenamefont
  {C\^andido~Ribeiro},\ and\ \citenamefont {Galetti}}]{PhysRevLett.79.5218}%
  \BibitemOpen
  \bibfield  {author} {\bibinfo {author} {\bibfnamefont {L.~C.}\ \bibnamefont
  {Chamon}}, \bibinfo {author} {\bibfnamefont {D.}~\bibnamefont {Pereira}},
  \bibinfo {author} {\bibfnamefont {M.~S.}\ \bibnamefont {Hussein}}, \bibinfo
  {author} {\bibfnamefont {M.~A.}\ \bibnamefont {C\^andido~Ribeiro}}, \ and\
  \bibinfo {author} {\bibfnamefont {D.}~\bibnamefont {Galetti}},\ }\href
  {\doibase 10.1103/PhysRevLett.79.5218} {\bibfield  {journal} {\bibinfo
  {journal} {Phys. Rev. Lett.}\ }\textbf {\bibinfo {volume} {79}},\ \bibinfo
  {pages} {5218} (\bibinfo {year} {1997})}\BibitemShut {NoStop}%
\bibitem [{\citenamefont {Chamon}\ \emph {et~al.}(1998)\citenamefont {Chamon},
  \citenamefont {Pereira},\ and\ \citenamefont {Hussein}}]{PhysRevC.58.576}%
  \BibitemOpen
  \bibfield  {author} {\bibinfo {author} {\bibfnamefont {L.~C.}\ \bibnamefont
  {Chamon}}, \bibinfo {author} {\bibfnamefont {D.}~\bibnamefont {Pereira}}, \
  and\ \bibinfo {author} {\bibfnamefont {M.~S.}\ \bibnamefont {Hussein}},\
  }\href {\doibase 10.1103/PhysRevC.58.576} {\bibfield  {journal} {\bibinfo
  {journal} {Phys. Rev. C}\ }\textbf {\bibinfo {volume} {58}},\ \bibinfo
  {pages} {576} (\bibinfo {year} {1998})}\BibitemShut {NoStop}%
\bibitem [{\citenamefont {Feshbach}(1992)}]{Fesh92}%
  \BibitemOpen
  \bibfield  {author} {\bibinfo {author} {\bibfnamefont {H.}~\bibnamefont
  {Feshbach}},\ }\href@noop {} {\bibfield  {journal} {\bibinfo  {journal}
  {Theoretical Nuclear Physics, Wiley, New York, 1992}\ }\textbf {\bibinfo
  {volume} {1}},\ \bibinfo {pages} {1} (\bibinfo {year} {1992})}\BibitemShut
  {NoStop}%
\bibitem [{\citenamefont {Brandan}\ and\ \citenamefont
  {McVoy}(1997)}]{PhysRevC.55.1362}%
  \BibitemOpen
  \bibfield  {author} {\bibinfo {author} {\bibfnamefont {M.~E.}\ \bibnamefont
  {Brandan}}\ and\ \bibinfo {author} {\bibfnamefont {K.~W.}\ \bibnamefont
  {McVoy}},\ }\href {\doibase 10.1103/PhysRevC.55.1362} {\bibfield  {journal}
  {\bibinfo  {journal} {Phys. Rev. C}\ }\textbf {\bibinfo {volume} {55}},\
  \bibinfo {pages} {1362} (\bibinfo {year} {1997})}\BibitemShut {NoStop}%
\bibitem [{\citenamefont {Brandan}\ and\ \citenamefont
  {Satchler}(1997)}]{S0370-1573(96)00048-8}%
  \BibitemOpen
  \bibfield  {author} {\bibinfo {author} {\bibfnamefont {M.~E.}\ \bibnamefont
  {Brandan}}\ and\ \bibinfo {author} {\bibfnamefont {G.~R.}\ \bibnamefont
  {Satchler}},\ }\href {\doibase 10.1016/S0370-1573(96)00048-8} {\bibfield
  {journal} {\bibinfo  {journal} {Phys. Rep.}\ }\textbf {\bibinfo {volume}
  {285}},\ \bibinfo {pages} {143} (\bibinfo {year} {1997})}\BibitemShut
  {NoStop}%
\bibitem [{\citenamefont {Alvarez}\ \emph {et~al.}(2003)\citenamefont
  {Alvarez}, \citenamefont {Chamon}, \citenamefont {Hussein}, \citenamefont
  {Pereira}, \citenamefont {Gasques}, \citenamefont {Rossi},\ and\
  \citenamefont {Silva}}]{S0375-9474(03)01158-8}%
  \BibitemOpen
  \bibfield  {author} {\bibinfo {author} {\bibfnamefont {M.~A.~G.}\
  \bibnamefont {Alvarez}}, \bibinfo {author} {\bibfnamefont {L.~C.}\
  \bibnamefont {Chamon}}, \bibinfo {author} {\bibfnamefont {M.~S.}\
  \bibnamefont {Hussein}}, \bibinfo {author} {\bibfnamefont {D.}~\bibnamefont
  {Pereira}}, \bibinfo {author} {\bibfnamefont {L.~R.}\ \bibnamefont
  {Gasques}}, \bibinfo {author} {\bibfnamefont {E.~S.}\ \bibnamefont {Rossi}},
  \ and\ \bibinfo {author} {\bibfnamefont {C.~P.}\ \bibnamefont {Silva}},\
  }\href {\doibase 10.1016/S0375-9474(03)01158-8} {\bibfield  {journal}
  {\bibinfo  {journal} {Nucl. Phys. A}\ }\textbf {\bibinfo {volume} {723}},\
  \bibinfo {pages} {93} (\bibinfo {year} {2003})}\BibitemShut {NoStop}%
\bibitem [{\citenamefont {Zagatto}\ \emph {et~al.}(2017)\citenamefont
  {Zagatto}, \citenamefont {Lubian}, \citenamefont {Gasques}, \citenamefont
  {Alvarez}, \citenamefont {Chamon}, \citenamefont {Oliveira}, \citenamefont
  {Alc\'antara-N\'u\~nez}, \citenamefont {Medina}, \citenamefont {Scarduelli},
  \citenamefont {Freitas}, \citenamefont {Padron}, \citenamefont {Rossi},\ and\
  \citenamefont {Shorto}}]{PhysRevC.95.064614}%
  \BibitemOpen
  \bibfield  {author} {\bibinfo {author} {\bibfnamefont {V.~A.~B.}\
  \bibnamefont {Zagatto}}, \bibinfo {author} {\bibfnamefont {J.}~\bibnamefont
  {Lubian}}, \bibinfo {author} {\bibfnamefont {L.~R.}\ \bibnamefont {Gasques}},
  \bibinfo {author} {\bibfnamefont {M.~A.~G.}\ \bibnamefont {Alvarez}},
  \bibinfo {author} {\bibfnamefont {L.~C.}\ \bibnamefont {Chamon}}, \bibinfo
  {author} {\bibfnamefont {J.~R.~B.}\ \bibnamefont {Oliveira}}, \bibinfo
  {author} {\bibfnamefont {J.~A.}\ \bibnamefont {Alc\'antara-N\'u\~nez}},
  \bibinfo {author} {\bibfnamefont {N.~H.}\ \bibnamefont {Medina}}, \bibinfo
  {author} {\bibfnamefont {V.}~\bibnamefont {Scarduelli}}, \bibinfo {author}
  {\bibfnamefont {A.}~\bibnamefont {Freitas}}, \bibinfo {author} {\bibfnamefont
  {I.}~\bibnamefont {Padron}}, \bibinfo {author} {\bibfnamefont {E.~S.}\
  \bibnamefont {Rossi}}, \ and\ \bibinfo {author} {\bibfnamefont {J.~M.~B.}\
  \bibnamefont {Shorto}},\ }\href {\doibase 10.1103/PhysRevC.95.064614}
  {\bibfield  {journal} {\bibinfo  {journal} {Phys. Rev. C}\ }\textbf {\bibinfo
  {volume} {95}},\ \bibinfo {pages} {064614} (\bibinfo {year}
  {2017})}\BibitemShut {NoStop}%
\bibitem [{\citenamefont {Arazi}\ \emph {et~al.}(2018)\citenamefont {Arazi},
  \citenamefont {Casal}, \citenamefont {Rodr\'{\i}guez-Gallardo}, \citenamefont
  {Arias}, \citenamefont {Lichtenth\"aler~Filho}, \citenamefont {Abriola},
  \citenamefont {Capurro}, \citenamefont {Cardona}, \citenamefont {Carnelli},
  \citenamefont {de~Barbar\'a}, \citenamefont {Fern\'andez~Niello},
  \citenamefont {Figueira}, \citenamefont {Fimiani}, \citenamefont {Hojman},
  \citenamefont {Mart\'{\i}}, \citenamefont {Mart\'{\i}nez~Heimman},\ and\
  \citenamefont {Pacheco}}]{PhysRevC.97.044609}%
  \BibitemOpen
  \bibfield  {author} {\bibinfo {author} {\bibfnamefont {A.}~\bibnamefont
  {Arazi}}, \bibinfo {author} {\bibfnamefont {J.}~\bibnamefont {Casal}},
  \bibinfo {author} {\bibfnamefont {M.}~\bibnamefont
  {Rodr\'{\i}guez-Gallardo}}, \bibinfo {author} {\bibfnamefont {J.~M.}\
  \bibnamefont {Arias}}, \bibinfo {author} {\bibfnamefont {R.}~\bibnamefont
  {Lichtenth\"aler~Filho}}, \bibinfo {author} {\bibfnamefont {D.}~\bibnamefont
  {Abriola}}, \bibinfo {author} {\bibfnamefont {O.~A.}\ \bibnamefont
  {Capurro}}, \bibinfo {author} {\bibfnamefont {M.~A.}\ \bibnamefont
  {Cardona}}, \bibinfo {author} {\bibfnamefont {P.~F.~F.}\ \bibnamefont
  {Carnelli}}, \bibinfo {author} {\bibfnamefont {E.}~\bibnamefont
  {de~Barbar\'a}}, \bibinfo {author} {\bibfnamefont {J.}~\bibnamefont
  {Fern\'andez~Niello}}, \bibinfo {author} {\bibfnamefont {J.~M.}\ \bibnamefont
  {Figueira}}, \bibinfo {author} {\bibfnamefont {L.}~\bibnamefont {Fimiani}},
  \bibinfo {author} {\bibfnamefont {D.}~\bibnamefont {Hojman}}, \bibinfo
  {author} {\bibfnamefont {G.~V.}\ \bibnamefont {Mart\'{\i}}}, \bibinfo
  {author} {\bibfnamefont {D.}~\bibnamefont {Mart\'{\i}nez~Heimman}}, \ and\
  \bibinfo {author} {\bibfnamefont {A.~J.}\ \bibnamefont {Pacheco}},\ }\href
  {\doibase 10.1103/PhysRevC.97.044609} {\bibfield  {journal} {\bibinfo
  {journal} {Phys. Rev. C}\ }\textbf {\bibinfo {volume} {97}},\ \bibinfo
  {pages} {044609} (\bibinfo {year} {2018})}\BibitemShut {NoStop}%
\bibitem [{\citenamefont {Gasques}\ \emph {et~al.}(2018)\citenamefont
  {Gasques}, \citenamefont {Freitas}, \citenamefont {Chamon}, \citenamefont
  {Oliveira}, \citenamefont {Medina}, \citenamefont {Scarduelli}, \citenamefont
  {Rossi}, \citenamefont {Alvarez}, \citenamefont {Zagatto}, \citenamefont
  {Lubian}, \citenamefont {Nobre}, \citenamefont {Padron},\ and\ \citenamefont
  {Carlson}}]{PhysRevC.97.034629}%
  \BibitemOpen
  \bibfield  {author} {\bibinfo {author} {\bibfnamefont {L.~R.}\ \bibnamefont
  {Gasques}}, \bibinfo {author} {\bibfnamefont {A.~S.}\ \bibnamefont
  {Freitas}}, \bibinfo {author} {\bibfnamefont {L.~C.}\ \bibnamefont {Chamon}},
  \bibinfo {author} {\bibfnamefont {J.~R.~B.}\ \bibnamefont {Oliveira}},
  \bibinfo {author} {\bibfnamefont {N.~H.}\ \bibnamefont {Medina}}, \bibinfo
  {author} {\bibfnamefont {V.}~\bibnamefont {Scarduelli}}, \bibinfo {author}
  {\bibfnamefont {E.~S.}\ \bibnamefont {Rossi}}, \bibinfo {author}
  {\bibfnamefont {M.~A.~G.}\ \bibnamefont {Alvarez}}, \bibinfo {author}
  {\bibfnamefont {V.~A.~B.}\ \bibnamefont {Zagatto}}, \bibinfo {author}
  {\bibfnamefont {J.}~\bibnamefont {Lubian}}, \bibinfo {author} {\bibfnamefont
  {G.~P.~A.}\ \bibnamefont {Nobre}}, \bibinfo {author} {\bibfnamefont
  {I.}~\bibnamefont {Padron}}, \ and\ \bibinfo {author} {\bibfnamefont {B.~V.}\
  \bibnamefont {Carlson}},\ }\href {\doibase 10.1103/PhysRevC.97.034629}
  {\bibfield  {journal} {\bibinfo  {journal} {Phys. Rev. C}\ }\textbf {\bibinfo
  {volume} {97}},\ \bibinfo {pages} {034629} (\bibinfo {year}
  {2018})}\BibitemShut {NoStop}%
\bibitem [{\citenamefont {Alvarez}\ \emph {et~al.}(2018)\citenamefont
  {Alvarez}, \citenamefont {Rodr\'{\i}guez-Gallardo}, \citenamefont {Gasques},
  \citenamefont {Chamon}, \citenamefont {Oliveira}, \citenamefont {Scarduelli},
  \citenamefont {Freitas}, \citenamefont {Rossi}, \citenamefont {Zagatto},
  \citenamefont {Rangel}, \citenamefont {Lubian},\ and\ \citenamefont
  {Padron}}]{PhysRevC.98.024621}%
  \BibitemOpen
  \bibfield  {author} {\bibinfo {author} {\bibfnamefont {M.~A.~G.}\
  \bibnamefont {Alvarez}}, \bibinfo {author} {\bibfnamefont {M.}~\bibnamefont
  {Rodr\'{\i}guez-Gallardo}}, \bibinfo {author} {\bibfnamefont {L.~R.}\
  \bibnamefont {Gasques}}, \bibinfo {author} {\bibfnamefont {L.~C.}\
  \bibnamefont {Chamon}}, \bibinfo {author} {\bibfnamefont {J.~R.~B.}\
  \bibnamefont {Oliveira}}, \bibinfo {author} {\bibfnamefont {V.}~\bibnamefont
  {Scarduelli}}, \bibinfo {author} {\bibfnamefont {A.~S.}\ \bibnamefont
  {Freitas}}, \bibinfo {author} {\bibfnamefont {E.~S.}\ \bibnamefont {Rossi}},
  \bibinfo {author} {\bibfnamefont {V.~A.~B.}\ \bibnamefont {Zagatto}},
  \bibinfo {author} {\bibfnamefont {J.}~\bibnamefont {Rangel}}, \bibinfo
  {author} {\bibfnamefont {J.}~\bibnamefont {Lubian}}, \ and\ \bibinfo {author}
  {\bibfnamefont {I.}~\bibnamefont {Padron}},\ }\href {\doibase
  10.1103/PhysRevC.98.024621} {\bibfield  {journal} {\bibinfo  {journal} {Phys.
  Rev. C}\ }\textbf {\bibinfo {volume} {98}},\ \bibinfo {pages} {024621}
  (\bibinfo {year} {2018})}\BibitemShut {NoStop}%
\bibitem [{\citenamefont {Gasques}\ \emph {et~al.}(2020)\citenamefont
  {Gasques}, \citenamefont {Chamon}, \citenamefont {L\'epine-Szily},
  \citenamefont {Scarduelli}, \citenamefont {Zagatto}, \citenamefont {Abriola},
  \citenamefont {Arazi}, \citenamefont {Cardona}, \citenamefont {de~Barbar\'a},
  \citenamefont {de~Jes\'us}, \citenamefont {Hojman}, \citenamefont
  {Mart\'{\i}}, \citenamefont {Pacheco}, \citenamefont {Ramos~L\'opez},
  \citenamefont {Alvarez}, \citenamefont {Fern\'andez-Garc\'{\i}a},
  \citenamefont {Rodr\'{\i}guez-Gallardo}, \citenamefont {Cubero},
  \citenamefont {Uma\~na},\ and\ \citenamefont
  {Ach\'{\i}-Prado}}]{PhysRevC.101.044604}%
  \BibitemOpen
  \bibfield  {author} {\bibinfo {author} {\bibfnamefont {L.~R.}\ \bibnamefont
  {Gasques}}, \bibinfo {author} {\bibfnamefont {L.~C.}\ \bibnamefont {Chamon}},
  \bibinfo {author} {\bibfnamefont {A.}~\bibnamefont {L\'epine-Szily}},
  \bibinfo {author} {\bibfnamefont {V.}~\bibnamefont {Scarduelli}}, \bibinfo
  {author} {\bibfnamefont {V.~A.~B.}\ \bibnamefont {Zagatto}}, \bibinfo
  {author} {\bibfnamefont {D.}~\bibnamefont {Abriola}}, \bibinfo {author}
  {\bibfnamefont {A.}~\bibnamefont {Arazi}}, \bibinfo {author} {\bibfnamefont
  {M.~A.}\ \bibnamefont {Cardona}}, \bibinfo {author} {\bibfnamefont
  {E.}~\bibnamefont {de~Barbar\'a}}, \bibinfo {author} {\bibfnamefont
  {J.}~\bibnamefont {de~Jes\'us}}, \bibinfo {author} {\bibfnamefont
  {D.}~\bibnamefont {Hojman}}, \bibinfo {author} {\bibfnamefont {G.~V.}\
  \bibnamefont {Mart\'{\i}}}, \bibinfo {author} {\bibfnamefont {A.~J.}\
  \bibnamefont {Pacheco}}, \bibinfo {author} {\bibfnamefont {D.}~\bibnamefont
  {Ramos~L\'opez}}, \bibinfo {author} {\bibfnamefont {M.~A.~G.}\ \bibnamefont
  {Alvarez}}, \bibinfo {author} {\bibfnamefont {J.~P.}\ \bibnamefont
  {Fern\'andez-Garc\'{\i}a}}, \bibinfo {author} {\bibfnamefont
  {M.}~\bibnamefont {Rodr\'{\i}guez-Gallardo}}, \bibinfo {author}
  {\bibfnamefont {M.}~\bibnamefont {Cubero}}, \bibinfo {author} {\bibfnamefont
  {L.~F.}\ \bibnamefont {Uma\~na}}, \ and\ \bibinfo {author} {\bibfnamefont
  {S.}~\bibnamefont {Ach\'{\i}-Prado}},\ }\href {\doibase
  10.1103/PhysRevC.101.044604} {\bibfield  {journal} {\bibinfo  {journal}
  {Phys. Rev. C}\ }\textbf {\bibinfo {volume} {101}},\ \bibinfo {pages}
  {044604} (\bibinfo {year} {2020})}\BibitemShut {NoStop}%
\bibitem [{\citenamefont {Aversa}\ \emph {et~al.}(2020)\citenamefont {Aversa},
  \citenamefont {Abriola}, \citenamefont {Alvarez}, \citenamefont {Arazi},
  \citenamefont {Cardona}, \citenamefont {Chamon}, \citenamefont
  {de~Barbar\'a}, \citenamefont {de~Jes\'us}, \citenamefont
  {Fern\'andez-Garc\'{\i}a}, \citenamefont {Gasques}, \citenamefont {Hojman},
  \citenamefont {L\'epine-Szily}, \citenamefont {Mart\'{\i}}, \citenamefont
  {Pacheco}, \citenamefont {Scarduelli},\ and\ \citenamefont
  {Zagatto}}]{PhysRevC.101.044601}%
  \BibitemOpen
  \bibfield  {author} {\bibinfo {author} {\bibfnamefont {M.}~\bibnamefont
  {Aversa}}, \bibinfo {author} {\bibfnamefont {D.}~\bibnamefont {Abriola}},
  \bibinfo {author} {\bibfnamefont {M.~A.~G.}\ \bibnamefont {Alvarez}},
  \bibinfo {author} {\bibfnamefont {A.}~\bibnamefont {Arazi}}, \bibinfo
  {author} {\bibfnamefont {M.~A.}\ \bibnamefont {Cardona}}, \bibinfo {author}
  {\bibfnamefont {L.~C.}\ \bibnamefont {Chamon}}, \bibinfo {author}
  {\bibfnamefont {E.}~\bibnamefont {de~Barbar\'a}}, \bibinfo {author}
  {\bibfnamefont {J.}~\bibnamefont {de~Jes\'us}}, \bibinfo {author}
  {\bibfnamefont {J.~P.}\ \bibnamefont {Fern\'andez-Garc\'{\i}a}}, \bibinfo
  {author} {\bibfnamefont {L.~R.}\ \bibnamefont {Gasques}}, \bibinfo {author}
  {\bibfnamefont {D.}~\bibnamefont {Hojman}}, \bibinfo {author} {\bibfnamefont
  {A.}~\bibnamefont {L\'epine-Szily}}, \bibinfo {author} {\bibfnamefont
  {G.~V.}\ \bibnamefont {Mart\'{\i}}}, \bibinfo {author} {\bibfnamefont
  {A.~J.}\ \bibnamefont {Pacheco}}, \bibinfo {author} {\bibfnamefont
  {V.}~\bibnamefont {Scarduelli}}, \ and\ \bibinfo {author} {\bibfnamefont
  {V.~A.~B.}\ \bibnamefont {Zagatto}},\ }\href {\doibase
  10.1103/PhysRevC.101.044601} {\bibfield  {journal} {\bibinfo  {journal}
  {Phys. Rev. C}\ }\textbf {\bibinfo {volume} {101}},\ \bibinfo {pages}
  {044601} (\bibinfo {year} {2020})}\BibitemShut {NoStop}%
\bibitem [{\citenamefont {Alvarez}\ \emph {et~al.}(2019)\citenamefont
  {Alvarez}, \citenamefont {Fern\'andez-Garc\'{\i}a}, \citenamefont
  {Le\'on-Garc\'{\i}a}, \citenamefont {Rodr\'{\i}guez-Gallardo}, \citenamefont
  {Gasques}, \citenamefont {Chamon}, \citenamefont {Zagatto}, \citenamefont
  {L\'epine-Szily}, \citenamefont {Oliveira}, \citenamefont {Scarduelli},
  \citenamefont {Carlson}, \citenamefont {Casal}, \citenamefont {Arazi},
  \citenamefont {Torres},\ and\ \citenamefont {Ramirez}}]{PhysRevC.100.064602}%
  \BibitemOpen
  \bibfield  {author} {\bibinfo {author} {\bibfnamefont {M.~A.~G.}\
  \bibnamefont {Alvarez}}, \bibinfo {author} {\bibfnamefont {J.~P.}\
  \bibnamefont {Fern\'andez-Garc\'{\i}a}}, \bibinfo {author} {\bibfnamefont
  {J.~L.}\ \bibnamefont {Le\'on-Garc\'{\i}a}}, \bibinfo {author} {\bibfnamefont
  {M.}~\bibnamefont {Rodr\'{\i}guez-Gallardo}}, \bibinfo {author}
  {\bibfnamefont {L.~R.}\ \bibnamefont {Gasques}}, \bibinfo {author}
  {\bibfnamefont {L.~C.}\ \bibnamefont {Chamon}}, \bibinfo {author}
  {\bibfnamefont {V.~A.~B.}\ \bibnamefont {Zagatto}}, \bibinfo {author}
  {\bibfnamefont {A.}~\bibnamefont {L\'epine-Szily}}, \bibinfo {author}
  {\bibfnamefont {J.~R.~B.}\ \bibnamefont {Oliveira}}, \bibinfo {author}
  {\bibfnamefont {V.}~\bibnamefont {Scarduelli}}, \bibinfo {author}
  {\bibfnamefont {B.~V.}\ \bibnamefont {Carlson}}, \bibinfo {author}
  {\bibfnamefont {J.}~\bibnamefont {Casal}}, \bibinfo {author} {\bibfnamefont
  {A.}~\bibnamefont {Arazi}}, \bibinfo {author} {\bibfnamefont {D.~A.}\
  \bibnamefont {Torres}}, \ and\ \bibinfo {author} {\bibfnamefont
  {F.}~\bibnamefont {Ramirez}},\ }\href {\doibase 10.1103/PhysRevC.100.064602}
  {\bibfield  {journal} {\bibinfo  {journal} {Phys. Rev. C}\ }\textbf {\bibinfo
  {volume} {100}},\ \bibinfo {pages} {064602} (\bibinfo {year}
  {2019})}\BibitemShut {NoStop}%
\bibitem [{\citenamefont {Nakamura}\ \emph {et~al.}(2006)\citenamefont
  {Nakamura}, \citenamefont {Vinodkumar}, \citenamefont {Sugimoto},
  \citenamefont {Aoi}, \citenamefont {Baba}, \citenamefont {Bazin},
  \citenamefont {Fukuda}, \citenamefont {Gomi}, \citenamefont {Hasegawa},
  \citenamefont {Imai}, \citenamefont {Ishihara}, \citenamefont {Kobayashi},
  \citenamefont {Kondo}, \citenamefont {Kubo}, \citenamefont {Miura},
  \citenamefont {Motobayashi}, \citenamefont {Otsu}, \citenamefont {Saito},
  \citenamefont {Sakurai}, \citenamefont {Shimoura}, \citenamefont {Watanabe},
  \citenamefont {Watanabe}, \citenamefont {Yakushiji}, \citenamefont
  {Yanagisawa},\ and\ \citenamefont {Yoneda}}]{PhysRevLett.96.252502}%
  \BibitemOpen
  \bibfield  {author} {\bibinfo {author} {\bibfnamefont {T.}~\bibnamefont
  {Nakamura}}, \bibinfo {author} {\bibfnamefont {A.~M.}\ \bibnamefont
  {Vinodkumar}}, \bibinfo {author} {\bibfnamefont {T.}~\bibnamefont
  {Sugimoto}}, \bibinfo {author} {\bibfnamefont {N.}~\bibnamefont {Aoi}},
  \bibinfo {author} {\bibfnamefont {H.}~\bibnamefont {Baba}}, \bibinfo {author}
  {\bibfnamefont {D.}~\bibnamefont {Bazin}}, \bibinfo {author} {\bibfnamefont
  {N.}~\bibnamefont {Fukuda}}, \bibinfo {author} {\bibfnamefont
  {T.}~\bibnamefont {Gomi}}, \bibinfo {author} {\bibfnamefont {H.}~\bibnamefont
  {Hasegawa}}, \bibinfo {author} {\bibfnamefont {N.}~\bibnamefont {Imai}},
  \bibinfo {author} {\bibfnamefont {M.}~\bibnamefont {Ishihara}}, \bibinfo
  {author} {\bibfnamefont {T.}~\bibnamefont {Kobayashi}}, \bibinfo {author}
  {\bibfnamefont {Y.}~\bibnamefont {Kondo}}, \bibinfo {author} {\bibfnamefont
  {T.}~\bibnamefont {Kubo}}, \bibinfo {author} {\bibfnamefont {M.}~\bibnamefont
  {Miura}}, \bibinfo {author} {\bibfnamefont {T.}~\bibnamefont {Motobayashi}},
  \bibinfo {author} {\bibfnamefont {H.}~\bibnamefont {Otsu}}, \bibinfo {author}
  {\bibfnamefont {A.}~\bibnamefont {Saito}}, \bibinfo {author} {\bibfnamefont
  {H.}~\bibnamefont {Sakurai}}, \bibinfo {author} {\bibfnamefont
  {S.}~\bibnamefont {Shimoura}}, \bibinfo {author} {\bibfnamefont
  {K.}~\bibnamefont {Watanabe}}, \bibinfo {author} {\bibfnamefont {Y.~X.}\
  \bibnamefont {Watanabe}}, \bibinfo {author} {\bibfnamefont {T.}~\bibnamefont
  {Yakushiji}}, \bibinfo {author} {\bibfnamefont {Y.}~\bibnamefont
  {Yanagisawa}}, \ and\ \bibinfo {author} {\bibfnamefont {K.}~\bibnamefont
  {Yoneda}},\ }\href {\doibase 10.1103/PhysRevLett.96.252502} {\bibfield
  {journal} {\bibinfo  {journal} {Phys. Rev. Lett.}\ }\textbf {\bibinfo
  {volume} {96}},\ \bibinfo {pages} {252502} (\bibinfo {year}
  {2006})}\BibitemShut {NoStop}%
\bibitem [{\citenamefont {Fukuda}\ \emph {et~al.}(2004)\citenamefont {Fukuda},
  \citenamefont {Nakamura}, \citenamefont {Aoi}, \citenamefont {Imai},
  \citenamefont {Ishihara}, \citenamefont {Kobayashi}, \citenamefont {Iwasaki},
  \citenamefont {Kubo}, \citenamefont {Mengoni}, \citenamefont {Notani},
  \citenamefont {Otsu}, \citenamefont {Sakurai}, \citenamefont {Shimoura},
  \citenamefont {Teranishi}, \citenamefont {Watanabe},\ and\ \citenamefont
  {Yoneda}}]{PhysRevC.70.054606}%
  \BibitemOpen
  \bibfield  {author} {\bibinfo {author} {\bibfnamefont {N.}~\bibnamefont
  {Fukuda}}, \bibinfo {author} {\bibfnamefont {T.}~\bibnamefont {Nakamura}},
  \bibinfo {author} {\bibfnamefont {N.}~\bibnamefont {Aoi}}, \bibinfo {author}
  {\bibfnamefont {N.}~\bibnamefont {Imai}}, \bibinfo {author} {\bibfnamefont
  {M.}~\bibnamefont {Ishihara}}, \bibinfo {author} {\bibfnamefont
  {T.}~\bibnamefont {Kobayashi}}, \bibinfo {author} {\bibfnamefont
  {H.}~\bibnamefont {Iwasaki}}, \bibinfo {author} {\bibfnamefont
  {T.}~\bibnamefont {Kubo}}, \bibinfo {author} {\bibfnamefont {A.}~\bibnamefont
  {Mengoni}}, \bibinfo {author} {\bibfnamefont {M.}~\bibnamefont {Notani}},
  \bibinfo {author} {\bibfnamefont {H.}~\bibnamefont {Otsu}}, \bibinfo {author}
  {\bibfnamefont {H.}~\bibnamefont {Sakurai}}, \bibinfo {author} {\bibfnamefont
  {S.}~\bibnamefont {Shimoura}}, \bibinfo {author} {\bibfnamefont
  {T.}~\bibnamefont {Teranishi}}, \bibinfo {author} {\bibfnamefont {Y.~X.}\
  \bibnamefont {Watanabe}}, \ and\ \bibinfo {author} {\bibfnamefont
  {K.}~\bibnamefont {Yoneda}},\ }\href {\doibase 10.1103/PhysRevC.70.054606}
  {\bibfield  {journal} {\bibinfo  {journal} {Phys. Rev. C}\ }\textbf {\bibinfo
  {volume} {70}},\ \bibinfo {pages} {054606} (\bibinfo {year}
  {2004})}\BibitemShut {NoStop}%
\bibitem [{\citenamefont {Casal}\ \emph {et~al.}(2014)\citenamefont {Casal},
  \citenamefont {Rodr\'{\i}guez-Gallardo}, \citenamefont {Arias},\ and\
  \citenamefont {Thompson}}]{PhysRevC.90.044304}%
  \BibitemOpen
  \bibfield  {author} {\bibinfo {author} {\bibfnamefont {J.}~\bibnamefont
  {Casal}}, \bibinfo {author} {\bibfnamefont {M.}~\bibnamefont
  {Rodr\'{\i}guez-Gallardo}}, \bibinfo {author} {\bibfnamefont {J.~M.}\
  \bibnamefont {Arias}}, \ and\ \bibinfo {author} {\bibfnamefont {I.~J.}\
  \bibnamefont {Thompson}},\ }\href {\doibase 10.1103/PhysRevC.90.044304}
  {\bibfield  {journal} {\bibinfo  {journal} {Phys. Rev. C}\ }\textbf {\bibinfo
  {volume} {90}},\ \bibinfo {pages} {044304} (\bibinfo {year}
  {2014})}\BibitemShut {NoStop}%
\bibitem [{\citenamefont {Alhassid}\ \emph {et~al.}(1982)\citenamefont
  {Alhassid}, \citenamefont {Gai},\ and\ \citenamefont
  {Bertsch}}]{PhysRevLett.49.1482}%
  \BibitemOpen
  \bibfield  {author} {\bibinfo {author} {\bibfnamefont {Y.}~\bibnamefont
  {Alhassid}}, \bibinfo {author} {\bibfnamefont {M.}~\bibnamefont {Gai}}, \
  and\ \bibinfo {author} {\bibfnamefont {G.~F.}\ \bibnamefont {Bertsch}},\
  }\href {\doibase 10.1103/PhysRevLett.49.1482} {\bibfield  {journal} {\bibinfo
   {journal} {Phys. Rev. Lett.}\ }\textbf {\bibinfo {volume} {49}},\ \bibinfo
  {pages} {1482} (\bibinfo {year} {1982})}\BibitemShut {NoStop}%
\bibitem [{\citenamefont {Bertulani}\ \emph {et~al.}(1991)\citenamefont
  {Bertulani}, \citenamefont {Baur},\ and\ \citenamefont
  {Hussein}}]{0375-9474(91)90442-9}%
  \BibitemOpen
  \bibfield  {author} {\bibinfo {author} {\bibfnamefont {C.}~\bibnamefont
  {Bertulani}}, \bibinfo {author} {\bibfnamefont {G.}~\bibnamefont {Baur}}, \
  and\ \bibinfo {author} {\bibfnamefont {M.~S.}\ \bibnamefont {Hussein}},\
  }\href {\doibase 10.1016/0375-9474(91)90442-9} {\bibfield  {journal}
  {\bibinfo  {journal} {Nucl. Phys. A}\ }\textbf {\bibinfo {volume} {526}},\
  \bibinfo {pages} {751} (\bibinfo {year} {1991})}\BibitemShut {NoStop}%
\bibitem [{\citenamefont {Yahiro}\ \emph {et~al.}(1986)\citenamefont {Yahiro},
  \citenamefont {Iseri}, \citenamefont {Kameyama}, \citenamefont {Kamimura},\
  and\ \citenamefont {Kawai}}]{10.1143/PTPS.89.32}%
  \BibitemOpen
  \bibfield  {author} {\bibinfo {author} {\bibfnamefont {M.}~\bibnamefont
  {Yahiro}}, \bibinfo {author} {\bibfnamefont {Y.}~\bibnamefont {Iseri}},
  \bibinfo {author} {\bibfnamefont {H.}~\bibnamefont {Kameyama}}, \bibinfo
  {author} {\bibfnamefont {M.}~\bibnamefont {Kamimura}}, \ and\ \bibinfo
  {author} {\bibfnamefont {M.}~\bibnamefont {Kawai}},\ }\href {\doibase
  10.1143/PTPS.89.32} {\bibfield  {journal} {\bibinfo  {journal} {Progress of
  Theoretical Physics Supplement}\ }\textbf {\bibinfo {volume} {89}},\ \bibinfo
  {pages} {32} (\bibinfo {year} {1986})}\BibitemShut {NoStop}%
\bibitem [{\citenamefont {Austern}\ \emph {et~al.}(1987)\citenamefont
  {Austern}, \citenamefont {Iseri}, \citenamefont {Kamimura}, \citenamefont
  {M.}, \citenamefont {Rawitscher},\ and\ \citenamefont
  {Yahiro}}]{0370-1573(87)90094-9}%
  \BibitemOpen
  \bibfield  {author} {\bibinfo {author} {\bibfnamefont {N.}~\bibnamefont
  {Austern}}, \bibinfo {author} {\bibfnamefont {Y.}~\bibnamefont {Iseri}},
  \bibinfo {author} {\bibfnamefont {M.}~\bibnamefont {Kamimura}}, \bibinfo
  {author} {\bibfnamefont {K.}~\bibnamefont {M.}}, \bibinfo {author}
  {\bibfnamefont {G.}~\bibnamefont {Rawitscher}}, \ and\ \bibinfo {author}
  {\bibfnamefont {M.}~\bibnamefont {Yahiro}},\ }\href {\doibase
  10.1016/0370-1573(87)90094-9} {\bibfield  {journal} {\bibinfo  {journal}
  {Phys. Rep.}\ }\textbf {\bibinfo {volume} {154}},\ \bibinfo {pages} {125}
  (\bibinfo {year} {1987})}\BibitemShut {NoStop}%
\bibitem [{\citenamefont {Cubero}\ \emph {et~al.}(2012)\citenamefont {Cubero},
  \citenamefont {Fern\'andez-Garc\'{\i}a}, \citenamefont
  {Rodr\'{\i}guez-Gallardo}, \citenamefont {Acosta}, \citenamefont {Alcorta},
  \citenamefont {Alvarez}, \citenamefont {Borge}, \citenamefont {Buchmann},
  \citenamefont {Diget}, \citenamefont {Falou}, \citenamefont {Fulton},
  \citenamefont {Fynbo}, \citenamefont {Galaviz}, \citenamefont
  {G\'omez-Camacho}, \citenamefont {Kanungo}, \citenamefont {Lay},
  \citenamefont {Madurga}, \citenamefont {Martel}, \citenamefont {Moro},
  \citenamefont {Mukha}, \citenamefont {Nilsson}, \citenamefont
  {S\'anchez-Ben\'{\i}tez}, \citenamefont {Shotter}, \citenamefont {Tengblad},\
  and\ \citenamefont {Walden}}]{PhysRevLett.109.262701}%
  \BibitemOpen
  \bibfield  {author} {\bibinfo {author} {\bibfnamefont {M.}~\bibnamefont
  {Cubero}}, \bibinfo {author} {\bibfnamefont {J.~P.}\ \bibnamefont
  {Fern\'andez-Garc\'{\i}a}}, \bibinfo {author} {\bibfnamefont
  {M.}~\bibnamefont {Rodr\'{\i}guez-Gallardo}}, \bibinfo {author}
  {\bibfnamefont {L.}~\bibnamefont {Acosta}}, \bibinfo {author} {\bibfnamefont
  {M.}~\bibnamefont {Alcorta}}, \bibinfo {author} {\bibfnamefont {M.~A.~G.}\
  \bibnamefont {Alvarez}}, \bibinfo {author} {\bibfnamefont {M.~J.~G.}\
  \bibnamefont {Borge}}, \bibinfo {author} {\bibfnamefont {L.}~\bibnamefont
  {Buchmann}}, \bibinfo {author} {\bibfnamefont {C.~A.}\ \bibnamefont {Diget}},
  \bibinfo {author} {\bibfnamefont {H.~A.}\ \bibnamefont {Falou}}, \bibinfo
  {author} {\bibfnamefont {B.~R.}\ \bibnamefont {Fulton}}, \bibinfo {author}
  {\bibfnamefont {H.~O.~U.}\ \bibnamefont {Fynbo}}, \bibinfo {author}
  {\bibfnamefont {D.}~\bibnamefont {Galaviz}}, \bibinfo {author} {\bibfnamefont
  {J.}~\bibnamefont {G\'omez-Camacho}}, \bibinfo {author} {\bibfnamefont
  {R.}~\bibnamefont {Kanungo}}, \bibinfo {author} {\bibfnamefont {J.~A.}\
  \bibnamefont {Lay}}, \bibinfo {author} {\bibfnamefont {M.}~\bibnamefont
  {Madurga}}, \bibinfo {author} {\bibfnamefont {I.}~\bibnamefont {Martel}},
  \bibinfo {author} {\bibfnamefont {A.~M.}\ \bibnamefont {Moro}}, \bibinfo
  {author} {\bibfnamefont {I.}~\bibnamefont {Mukha}}, \bibinfo {author}
  {\bibfnamefont {T.}~\bibnamefont {Nilsson}}, \bibinfo {author} {\bibfnamefont
  {A.~M.}\ \bibnamefont {S\'anchez-Ben\'{\i}tez}}, \bibinfo {author}
  {\bibfnamefont {A.}~\bibnamefont {Shotter}}, \bibinfo {author} {\bibfnamefont
  {O.}~\bibnamefont {Tengblad}}, \ and\ \bibinfo {author} {\bibfnamefont
  {P.}~\bibnamefont {Walden}},\ }\href {\doibase
  10.1103/PhysRevLett.109.262701} {\bibfield  {journal} {\bibinfo  {journal}
  {Phys. Rev. Lett.}\ }\textbf {\bibinfo {volume} {109}},\ \bibinfo {pages}
  {262701} (\bibinfo {year} {2012})}\BibitemShut {NoStop}%
\bibitem [{\citenamefont {de~Faria}\ \emph {et~al.}(2010)\citenamefont
  {de~Faria}, \citenamefont {Lichtenth\"aler}, \citenamefont {Pires},
  \citenamefont {Moro}, \citenamefont {L\'epine-Szily}, \citenamefont
  {Guimar\~aes}, \citenamefont {Mendes}, \citenamefont {Arazi}, \citenamefont
  {Rodr\'{\i}guez-Gallardo}, \citenamefont {Barioni}, \citenamefont {Morcelle},
  \citenamefont {Morais}, \citenamefont {Camargo}, \citenamefont {Alcantara
  Nu\~nez},\ and\ \citenamefont
  {Assun\ifmmode\mbox{\c{c}}\else\c{c}\fi{}\~{a}o}}]{PhysRevC.81.044605}%
  \BibitemOpen
  \bibfield  {author} {\bibinfo {author} {\bibfnamefont {P.~N.}\ \bibnamefont
  {de~Faria}}, \bibinfo {author} {\bibfnamefont {R.}~\bibnamefont
  {Lichtenth\"aler}}, \bibinfo {author} {\bibfnamefont {K.~C.~C.}\ \bibnamefont
  {Pires}}, \bibinfo {author} {\bibfnamefont {A.~M.}\ \bibnamefont {Moro}},
  \bibinfo {author} {\bibfnamefont {A.}~\bibnamefont {L\'epine-Szily}},
  \bibinfo {author} {\bibfnamefont {V.}~\bibnamefont {Guimar\~aes}}, \bibinfo
  {author} {\bibfnamefont {D.~R.}\ \bibnamefont {Mendes}}, \bibinfo {author}
  {\bibfnamefont {A.}~\bibnamefont {Arazi}}, \bibinfo {author} {\bibfnamefont
  {M.}~\bibnamefont {Rodr\'{\i}guez-Gallardo}}, \bibinfo {author}
  {\bibfnamefont {A.}~\bibnamefont {Barioni}}, \bibinfo {author} {\bibfnamefont
  {V.}~\bibnamefont {Morcelle}}, \bibinfo {author} {\bibfnamefont {M.~C.}\
  \bibnamefont {Morais}}, \bibinfo {author} {\bibfnamefont {O.}~\bibnamefont
  {Camargo}}, \bibinfo {author} {\bibfnamefont {J.}~\bibnamefont {Alcantara
  Nu\~nez}}, \ and\ \bibinfo {author} {\bibfnamefont {M.}~\bibnamefont
  {Assun\ifmmode\mbox{\c{c}}\else\c{c}\fi{}\~{a}o}},\ }\href {\doibase
  10.1103/PhysRevC.81.044605} {\bibfield  {journal} {\bibinfo  {journal} {Phys.
  Rev. C}\ }\textbf {\bibinfo {volume} {81}},\ \bibinfo {pages} {044605}
  (\bibinfo {year} {2010})}\BibitemShut {NoStop}%
\bibitem [{\citenamefont {Casal}\ \emph {et~al.}(2015)\citenamefont {Casal},
  \citenamefont {Rodr\'{\i}guez-Gallardo},\ and\ \citenamefont
  {Arias}}]{PhysRevC.92.054611}%
  \BibitemOpen
  \bibfield  {author} {\bibinfo {author} {\bibfnamefont {J.}~\bibnamefont
  {Casal}}, \bibinfo {author} {\bibfnamefont {M.}~\bibnamefont
  {Rodr\'{\i}guez-Gallardo}}, \ and\ \bibinfo {author} {\bibfnamefont {J.~M.}\
  \bibnamefont {Arias}},\ }\href {\doibase 10.1103/PhysRevC.92.054611}
  {\bibfield  {journal} {\bibinfo  {journal} {Phys. Rev. C}\ }\textbf {\bibinfo
  {volume} {92}},\ \bibinfo {pages} {054611} (\bibinfo {year}
  {2015})}\BibitemShut {NoStop}%
\bibitem [{\citenamefont {Rodr\'{\i}guez-Gallardo}\ \emph
  {et~al.}(2009)\citenamefont {Rodr\'{\i}guez-Gallardo}, \citenamefont {Arias},
  \citenamefont {G\'omez-Camacho}, \citenamefont {Moro}, \citenamefont
  {Thompson},\ and\ \citenamefont {Tostevin}}]{PhysRevC.80.051601}%
  \BibitemOpen
  \bibfield  {author} {\bibinfo {author} {\bibfnamefont {M.}~\bibnamefont
  {Rodr\'{\i}guez-Gallardo}}, \bibinfo {author} {\bibfnamefont {J.~M.}\
  \bibnamefont {Arias}}, \bibinfo {author} {\bibfnamefont {J.}~\bibnamefont
  {G\'omez-Camacho}}, \bibinfo {author} {\bibfnamefont {A.~M.}\ \bibnamefont
  {Moro}}, \bibinfo {author} {\bibfnamefont {I.~J.}\ \bibnamefont {Thompson}},
  \ and\ \bibinfo {author} {\bibfnamefont {J.~A.}\ \bibnamefont {Tostevin}},\
  }\href {\doibase 10.1103/PhysRevC.80.051601} {\bibfield  {journal} {\bibinfo
  {journal} {Phys. Rev. C}\ }\textbf {\bibinfo {volume} {80}},\ \bibinfo
  {pages} {051601(R)} (\bibinfo {year} {2009})}\BibitemShut {NoStop}%
\bibitem [{\citenamefont {Casal}\ \emph {et~al.}(2013)\citenamefont {Casal},
  \citenamefont {Rodr\'{\i}guez-Gallardo},\ and\ \citenamefont
  {Arias}}]{PhysRevC.88.014327}%
  \BibitemOpen
  \bibfield  {author} {\bibinfo {author} {\bibfnamefont {J.}~\bibnamefont
  {Casal}}, \bibinfo {author} {\bibfnamefont {M.}~\bibnamefont
  {Rodr\'{\i}guez-Gallardo}}, \ and\ \bibinfo {author} {\bibfnamefont {J.~M.}\
  \bibnamefont {Arias}},\ }\href {\doibase 10.1103/PhysRevC.88.014327}
  {\bibfield  {journal} {\bibinfo  {journal} {Phys. Rev. C}\ }\textbf {\bibinfo
  {volume} {88}},\ \bibinfo {pages} {014327} (\bibinfo {year}
  {2013})}\BibitemShut {NoStop}%
\bibitem [{\citenamefont {Lay}\ \emph {et~al.}(2012)\citenamefont {Lay},
  \citenamefont {Moro}, \citenamefont {Arias},\ and\ \citenamefont
  {G\'omez-Camacho}}]{PhysRevC.85.054618}%
  \BibitemOpen
  \bibfield  {author} {\bibinfo {author} {\bibfnamefont {J.~A.}\ \bibnamefont
  {Lay}}, \bibinfo {author} {\bibfnamefont {A.~M.}\ \bibnamefont {Moro}},
  \bibinfo {author} {\bibfnamefont {J.~M.}\ \bibnamefont {Arias}}, \ and\
  \bibinfo {author} {\bibfnamefont {J.}~\bibnamefont {G\'omez-Camacho}},\
  }\href {\doibase 10.1103/PhysRevC.85.054618} {\bibfield  {journal} {\bibinfo
  {journal} {Phys. Rev. C}\ }\textbf {\bibinfo {volume} {85}},\ \bibinfo
  {pages} {054618} (\bibinfo {year} {2012})}\BibitemShut {NoStop}%
\bibitem [{\citenamefont {de~Diego}\ \emph {et~al.}(2014)\citenamefont
  {de~Diego}, \citenamefont {Arias}, \citenamefont {Lay},\ and\ \citenamefont
  {Moro}}]{PhysRevC.89.064609}%
  \BibitemOpen
  \bibfield  {author} {\bibinfo {author} {\bibfnamefont {R.}~\bibnamefont
  {de~Diego}}, \bibinfo {author} {\bibfnamefont {J.~M.}\ \bibnamefont {Arias}},
  \bibinfo {author} {\bibfnamefont {J.~A.}\ \bibnamefont {Lay}}, \ and\
  \bibinfo {author} {\bibfnamefont {A.~M.}\ \bibnamefont {Moro}},\ }\href
  {\doibase 10.1103/PhysRevC.89.064609} {\bibfield  {journal} {\bibinfo
  {journal} {Phys. Rev. C}\ }\textbf {\bibinfo {volume} {89}},\ \bibinfo
  {pages} {064609} (\bibinfo {year} {2014})}\BibitemShut {NoStop}%
\bibitem [{\citenamefont {Summers}\ \emph {et~al.}(2006)\citenamefont
  {Summers}, \citenamefont {Nunes},\ and\ \citenamefont
  {Thompson}}]{PhysRevC.74.014606}%
  \BibitemOpen
  \bibfield  {author} {\bibinfo {author} {\bibfnamefont {N.~C.}\ \bibnamefont
  {Summers}}, \bibinfo {author} {\bibfnamefont {F.~M.}\ \bibnamefont {Nunes}},
  \ and\ \bibinfo {author} {\bibfnamefont {I.~J.}\ \bibnamefont {Thompson}},\
  }\href {\doibase 10.1103/PhysRevC.74.014606} {\bibfield  {journal} {\bibinfo
  {journal} {Phys. Rev. C}\ }\textbf {\bibinfo {volume} {74}},\ \bibinfo
  {pages} {014606} (\bibinfo {year} {2006})}\BibitemShut {NoStop}%
\bibitem [{\citenamefont {Thompson}\ \emph {et~al.}(1989)\citenamefont
  {Thompson}, \citenamefont {Nagarajan}, \citenamefont {Lilley},\ and\
  \citenamefont {Smithson}}]{THOMPSON198984}%
  \BibitemOpen
  \bibfield  {author} {\bibinfo {author} {\bibfnamefont {I.}~\bibnamefont
  {Thompson}}, \bibinfo {author} {\bibfnamefont {M.}~\bibnamefont {Nagarajan}},
  \bibinfo {author} {\bibfnamefont {J.}~\bibnamefont {Lilley}}, \ and\ \bibinfo
  {author} {\bibfnamefont {M.}~\bibnamefont {Smithson}},\ }\href {\doibase
  https://doi.org/10.1016/0375-9474(89)90417-X} {\bibfield  {journal} {\bibinfo
   {journal} {Nuclear Physics A}\ }\textbf {\bibinfo {volume} {505}},\ \bibinfo
  {pages} {84} (\bibinfo {year} {1989})}\BibitemShut {NoStop}%
\bibitem [{\citenamefont {Yu}\ \emph {et~al.}(2010)\citenamefont {Yu},
  \citenamefont {Zhang}, \citenamefont {Jia}, \citenamefont {Zhang},
  \citenamefont {Ruan}, \citenamefont {Yang}, \citenamefont {Wu}, \citenamefont
  {Xu},\ and\ \citenamefont {Bai}}]{Yu075108}%
  \BibitemOpen
  \bibfield  {author} {\bibinfo {author} {\bibfnamefont {N.}~\bibnamefont
  {Yu}}, \bibinfo {author} {\bibfnamefont {H.~Q.}\ \bibnamefont {Zhang}},
  \bibinfo {author} {\bibfnamefont {H.~M.}\ \bibnamefont {Jia}}, \bibinfo
  {author} {\bibfnamefont {S.~T.}\ \bibnamefont {Zhang}}, \bibinfo {author}
  {\bibfnamefont {M.}~\bibnamefont {Ruan}}, \bibinfo {author} {\bibfnamefont
  {F.}~\bibnamefont {Yang}}, \bibinfo {author} {\bibfnamefont {Z.~D.}\
  \bibnamefont {Wu}}, \bibinfo {author} {\bibfnamefont {X.~X.}\ \bibnamefont
  {Xu}}, \ and\ \bibinfo {author} {\bibfnamefont {C.~L.}\ \bibnamefont {Bai}},\
  }\href {\doibase 10.1088/0954-3899/37/7/075108} {\bibfield  {journal}
  {\bibinfo  {journal} {J. of Phys. G: Nuclear and Particle Physics}\ }\textbf
  {\bibinfo {volume} {37}},\ \bibinfo {pages} {75} (\bibinfo {year}
  {2010})}\BibitemShut {NoStop}%
\bibitem [{\citenamefont {Mohr}\ \emph {et~al.}(2010)\citenamefont {Mohr},
  \citenamefont {de~Faria}, \citenamefont {Lichtenth\"aler}, \citenamefont
  {Pires}, \citenamefont {Guimar\~aes}, \citenamefont {L\'epine-Szily},
  \citenamefont {Mendes}, \citenamefont {Arazi}, \citenamefont {Barioni},
  \citenamefont {Morcelle},\ and\ \citenamefont {Morais}}]{PhysRevC.82.044606}%
  \BibitemOpen
  \bibfield  {author} {\bibinfo {author} {\bibfnamefont {P.}~\bibnamefont
  {Mohr}}, \bibinfo {author} {\bibfnamefont {P.~N.}\ \bibnamefont {de~Faria}},
  \bibinfo {author} {\bibfnamefont {R.}~\bibnamefont {Lichtenth\"aler}},
  \bibinfo {author} {\bibfnamefont {K.~C.~C.}\ \bibnamefont {Pires}}, \bibinfo
  {author} {\bibfnamefont {V.}~\bibnamefont {Guimar\~aes}}, \bibinfo {author}
  {\bibfnamefont {A.}~\bibnamefont {L\'epine-Szily}}, \bibinfo {author}
  {\bibfnamefont {D.~R.}\ \bibnamefont {Mendes}}, \bibinfo {author}
  {\bibfnamefont {A.}~\bibnamefont {Arazi}}, \bibinfo {author} {\bibfnamefont
  {A.}~\bibnamefont {Barioni}}, \bibinfo {author} {\bibfnamefont
  {V.}~\bibnamefont {Morcelle}}, \ and\ \bibinfo {author} {\bibfnamefont
  {M.~C.}\ \bibnamefont {Morais}},\ }\href {\doibase
  10.1103/PhysRevC.82.044606} {\bibfield  {journal} {\bibinfo  {journal} {Phys.
  Rev. C}\ }\textbf {\bibinfo {volume} {82}},\ \bibinfo {pages} {044606}
  (\bibinfo {year} {2010})}\BibitemShut {NoStop}%
\bibitem [{\citenamefont {Mackintosh}\ and\ \citenamefont
  {Keeley}(2009)}]{PhysRevC.79.014611}%
  \BibitemOpen
  \bibfield  {author} {\bibinfo {author} {\bibfnamefont {R.~S.}\ \bibnamefont
  {Mackintosh}}\ and\ \bibinfo {author} {\bibfnamefont {N.}~\bibnamefont
  {Keeley}},\ }\href {\doibase 10.1103/PhysRevC.79.014611} {\bibfield
  {journal} {\bibinfo  {journal} {Phys. Rev. C}\ }\textbf {\bibinfo {volume}
  {79}},\ \bibinfo {pages} {014611} (\bibinfo {year} {2009})}\BibitemShut
  {NoStop}%
\bibitem [{\citenamefont {Rusek}(2009)}]{RUSEK}%
  \BibitemOpen
  \bibfield  {author} {\bibinfo {author} {\bibfnamefont {K.}~\bibnamefont
  {Rusek}},\ }\href {\doibase 10.1140/epja/i2009-10838-x} {\bibfield  {journal}
  {\bibinfo  {journal} {European Physical Journal A}\ }\textbf {\bibinfo
  {volume} {41}},\ \bibinfo {pages} {399} (\bibinfo {year} {2009})}\BibitemShut
  {NoStop}%
\end{thebibliography}%

\end{document}